\documentclass[aps,pra,showpacs,twoside,twocolumn,10pt]{revtex4-1}
\usepackage[colorlinks=true, citecolor=red, urlcolor=blue ]{hyperref}
\usepackage{epsfig,newlfont,amssymb,amsfonts,amsmath,bm,subfigure,palatino,mathtools,amsthm,braket,soul,enumitem,color,graphics,graphicx,times,physics}

\usepackage[normalem]{ulem}
\usepackage{dsfont}
\usepackage{xcolor}
\newtheorem{prop}{Proposition}

\begin{document}
	
	\title{
		Measurement-based qudit quantum refrigerator with subspace cooling  
	}
	
	\author{Debkanta Ghosh, Tanoy Kanti Konar, Aditi Sen(De)}
	\affiliation{Harish-Chandra Research Institute, a CI of HBNI, Chhatnag Road, Jhunsi, Allahabad 211 019, India}

	\begin{abstract}
		
		We develop a method to transform a collection of higher-dimensional spin systems from the thermal state with a very high temperature of a local spin-\(s\) Hamiltonian to a low-lying energy eigenstate of the same. The procedure utilizes an auxiliary system, interactions between all systems, and appropriate projective measurements of arbitrary rank performed on the auxiliary system. We refer to this process as subspace cooling. The performance of the protocol is assessed by determining the fidelity of the target  state with the output one and the success probability of achieving the resulting state.  For this analysis, spin-\(s\) \(XXZ\) and bilinear biquadratic models are employed as the evolving Hamiltonian. We demonstrate that  in both scenarios, unit fidelity can be attained after a reasonable number of repeated measurements and a finite amount of evolution time when all the systems are aligned in an open chain, but it fails when the interactions between the spin follow the star configuration. We report that the success probability  increases with the rank of the projectors in the measurement for a fixed dimension and that for each dimension, there exists a range of interaction strength and evolution period for which the  fidelity gets maximized. Even when some subsystems are in contact with the thermal bath, the method proves to be resistant to decoherence.

	\end{abstract}
	
	\maketitle
	
	\section{Introduction}
	
	In the last two decades, the quest for thermodynamical laws at the microscopic regime \cite{gemmer2004} has resulted in the discovery of quantum thermal devices such as the quantum version of battery\cite{Alicki,modi2018,PoliniPRL,srijon2020,battery_review_rmp}, diode \cite{ordonez2017}, refrigerator \cite{linden2010,skrzypczyk2011, konar2021}, transistors \cite{karl2016,ahana_transistor}, and heat engines \cite{scovil_prl_heatengine} which are miniaturized versions of existing classical appliances.
	Additionally, these devices have  been  constructed  in laboratories employing  ultracold atoms \cite{Shapiro2012}, trapped ions \cite{haffner2008,rossnage2016}, superconducting \cite{karimi_prb_2016}, atoms in a cavity \cite{zhang_cavity_engine}, photonic \cite{Hardal2015Aug}, and nuclear magnetic resonance (NMR)  \cite{peterson2019,mahesh_quantumbattery}. 
	At the same time, successful manufacture of quantum technologies necessitates the development of appropriate states, which might be the ground state or the eigenstates of a specific Hamiltonian representing complex systems.  For instance, it has been demonstrated that creating quantum communication networks and measurement-based quantum computation \cite{raussendorf2001} requires sharing genuine multipartite entangled states \cite{horodecki2009} such as Greenberger Horne Zeilinger (GHZ) \cite{Greenberger2007}, W \cite{Durvidal2002} and cluster states \cite{raussendorf2001,mbqc_Raussendorf2003, mbqc1, mbqc2}.

	The system being investigated is always in contact with the environment, resulting in noisy states \cite{Petruccione}.  Hence, distilling pure states from noisy ones  becomes crucial for accomplishing quantum information tasks. In the literature, two routes of quantum distillation protocols are developed -- on the one hand, several copies of noisy entangled states are  purified to pure maximally entangled states via specific classes of quantum operations like local operations and classical communication --  (direct) distillation scheme \cite{Bennett96} -- direct distillation scheme; on the other hand, instead of directly measuring the mixed states to be purified, an indirect protocol uses auxiliary systems that are first attached to the target systems, followed by measurements and the process is typically repeated a large number of times to achieve the goal \cite{nakazato2003, nakazato2004, nakazato2008, bellomo2010, burgarth2007,burgarth2008}.  Since the latter process involves repeated measurement, it is called the Zeno-like measurement \cite{misra1977zeno,dega1974}.  Furthermore, the procedure might be referred to as refrigeration because the thermal state is converted into the ground state during the process  \cite{yan2021, yan2022, konar_measure_2022}. Following its inception, variety of protocols have been designed including the purification of multiple qubits, thereby preparing specific entangled states like maximally entangled two-qubit states \cite{nakazato2004} and GHZ states \cite{yan2023},   distilling vibrational states in trapped ions \cite{glans2004}, noncoherent control for purification \cite{romano2007}, purification of nuclear spin ensemble \cite{Greiner2017Apr},  cooling of resonator from thermal equilibrium state with high-temperature \cite{yan2021} and quantum batteries \cite{yan_battery_2023,chaki_battery_2023}.

	
	
	
	The purification strategies based on Zeno-like measurements are often presented for a collection of two-dimensional systems, with the target systems assumed to be pure. Our work departs from both of these limitations. Specifically, we formulate a qudit-based cooling or purifying scheme by exploiting  Zeno-like measurements.  Such consideration is motivated from the fact that higher-dimensional systems have been demonstrated to be more nonclassical than lower-dimensional systems \cite{Bellqudits, bell-brunner}, making them more favorable in the development of quantum technologies than their two-dimensional counterparts. Prominent examples include quantum key distribution \cite{QKDqudit1,gisin2002,scarani2010},  quantum computation \cite{Zhou_2003,Hosten_2005,hall2005cluster, joo_2007,Zhou_2009,Raussen_2017, Wang2020, Booth_2023,Marqversen2023,Imany2019,Sawant_2020,Low2020,Hrmo2023},  quantum thermal machines like quantum batteries \cite{ghosh2021}, quantum refrigerators \cite{correa2014,wang2015,silva2015,silva2016, usui2021, konar2022}.  In addition to theoretical advancements, qudit-based quantum information processing tasks have been realized in several physical systems such as trapped ions \cite{iontrap08,iontrap09,iontrap12}, photons \cite{pan2019,pan2021}, and  superconducting systems \cite{PRXQuantumsuper,ying23}.

	
	We exhibit that higher-dimensional systems can provide relaxation to the target systems to be cooled from  pure states to  equal mixtures of low-lying states of the initial Hamiltonian by measuring arbitrary rank projectors -- we refer to this concept as {\it subspace cooling}. This is particularly crucial since obtaining absolute zero temperature is difficult to achieve experimentally, hence nonvanishing temperature always results in mixtures of low-lying states.  
	To validate our protocol, each qudit is first prepared in a thermal state of a local Hamiltonian that overlaps with all excited states, while the auxiliary system is always assumed to be the desired target state.  Subspace cooling can be accomplished by interacting the qudits with the auxiliary system and then performing high-rank (greater than rank-1) projective measurements repeatedly.  To our knowledge, higher rank measurements have never been considered in Zeno-like measurement protocols to purify quantum states, hence only cooling to the ground state appears in the literature. To access the quality of cooling, we compute  Uhlmann fidelity \cite{uhlmann1976,jozsa1994,mendon2008,miszczak2008}   between the target  and the output state acquired  at the end of the protocol and  the success probability of generating the output. We report that, in the case of subspace cooling, proper tuning of the system parameters, evolution time and a reasonably high number of measurements performed can attain the unit fidelity of the target systems. Moreover, for a given dimension of each subsystem,  we show that the success probability for obtaining the desired measurement outcome  increases with the increase of the rank of the measurements,  demonstrating the significance of performing higher rank measurements during the purifying process. We illustrate the qudit-based protocol by utilizing  two different interacting Hamiltonians, namely spin -\(s\) \(XXZ\) \cite{dutta} and bilinear-biquadratic Heisenberg (BBH) \cite{Sutherland1975,Takhtajan1982,Babu82,Fath91,Fath93} models.

	We observe that for a fixed individual dimension of the target subsystems, refrigeration via a finite number of measurement is not achievable for a given interaction strength and evolution time between two measurements. Interestingly, we also notice that attaining maximal fidelity also depends on the geometry of the lattice according to which the evolution is carried out. In particular, in a star network, we are unable to determine the system parameters which lead to the unit fidelity provided the measurement is carried out on the central spin. Furthermore, we observe that the subspace cooling mechanism is robust against decoherence when some of the subsystems are in contact with the bath.



	
	The following is the organization of the paper.  We discuss the protocol for  refrigeration via rank-\(k\) measurements and introduce the idea of subspace cooling  in Sec. \ref{sec:framework}. Sec. \ref{sec:paradigmatic} presents the analytical results of the subspace cooling protocol for the paradigmatic interacting Hamiltonian and measurements. In Sec. \ref{sec:l_array}, the fidelity and the success probability are studied when  the target spin and an auxiliary system interact  in a linear chain configuration while in Sec. \ref{sec:lattice geometry}, we show that in a star configuration, the fidelity gets decreased due to the symmetry. The effects of decoherence on the  scheme are discussed in Sec. \ref{sec:evironment_effect} before the concluding remarks in Sec. \ref{sec:conclu}.

	\begin{figure*}
		\centering
		\includegraphics[width=\textwidth]{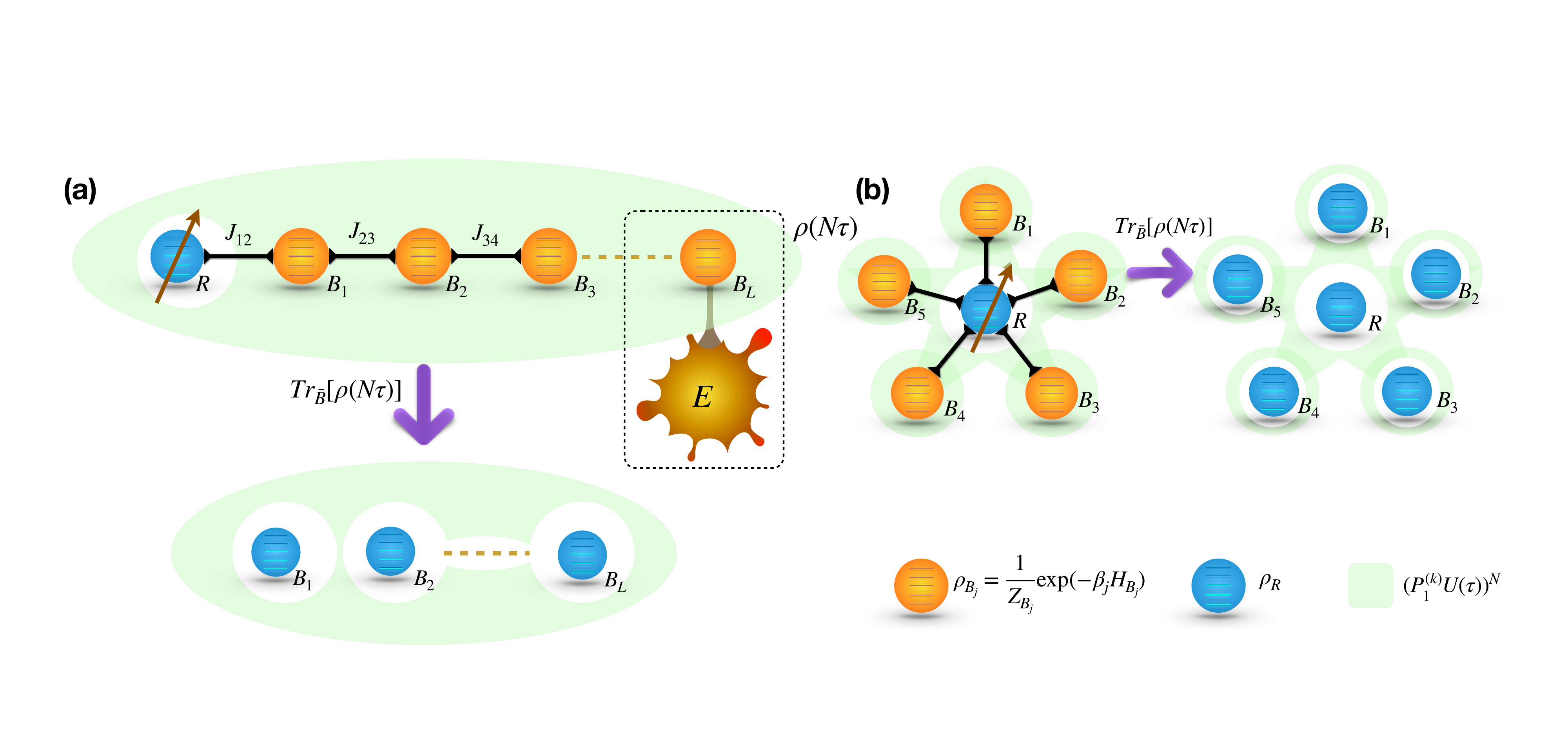}
		\caption{Schematic of a  subspace cooling process. The target (inaccessible) qudits (yellow ones), $B_{j}(j=1,2,\ldots,l)$, are prepared in the thermal state of $H_{B_j}$, \(\rho_{B_j}\) (yellow sphere) whereas the regulator qudit \(\rho_R\) (blue sphere) is prepared in the equal mixture of the low-lying energy eigenstates of \(H_{B_j}\). All the systems interact according to some Hamiltonian \(H_{int}\) for a time period \(\tau\) and  the rank-\(k\) projective measurement is performed on \(R\). The process is repeated for \(N\) times, i.e.,  \((P_1^{(k)}U(\tau))^N\). After a moderate number of repetitions, the resulting  target states become the same as the regular state.   We investigate our protocol when the interaction follows the lattice geometry (a) an one-dimensional array of $L$ spin-$s$ particles and (b) a star network of $L$ spin-$s$ particles though the protocol is independent of any lattice geometry. In (a), the last qudit is attached to a bath so that the environmental effect on the protocol can be studied. Note that $J_{ij}$ is the interaction strength between the sites $i$ and $j$.}
		\label{fig:schematic}
	\end{figure*}

	\section{Subspace cooling: Refrigeration using rank-\(k\) projectors}
	\label{sec:framework}
	Let us first describe the set-up of the quantum refrigerator. The refrigerator consists of $L$  number of spin-$s$ particles,  denoted as $B_j$ (where $j=1,2,\dots,L$), fixing the dimension  \(d=2s+1\), with \(s\) being the spin quantum number. It is controlled or regulated by a single  qudit, denoted as \(R\) since the measurement is performed on this qudit. Initially, all the target qudits, \(B_j\), are prepared in the thermal state of a  local Hamiltonian, $H_{B_j}$, represented as \(\rho_{in} = \otimes_{j=1}^L\rho_{B_j}=\otimes_{j=1}^L\frac{1}{Z_{B_j}}\exp(-\beta_j H_{B_j})\), where \(Z_{B_j}=\Tr[\exp(-\beta_j H_{B_j})]\), $\beta_j=\frac{1}{k_B T_j}$ is the inverse temperature of each qudit with temperature $T_j$ and $k_B$ is the Boltzmann constant.
	Hence, the subsystems have a finite overlap with all the eigenstates of \(H_{B_j}\) from lower energy to higher energy sectors. Our aim is to cool the individual subsystem by bringing an auxiliary qudit \(R\), $\rho_R$, so that the entire system is initially in a product form, \(\rho(0)=\rho_R\otimes\rho_{in}\). The refrigeration of individual qudits, \(B_j (\forall j)\), occurs by turning on the interactions between \(B_j\)s and, \(R\) along with the frequent projective measurement on $R$. Here, the cooling of $B_j$ means transforming the mixed initial states into ground (pure) states of the individual Hamiltonian, $H_{B_j}$ \cite{nakazato2003, nakazato2004, nakazato2008, bellomo2010}. Although in case of qubits, measurements on controlled qubits can only purify maximally mixed states to pure states with a non-vanishing probability. We will exhibit that in the case of higher dimensional system, more general framework of cooling can be developed. Specifically, the goal is to induce the transformation such that \(\rho(0)\to \rho(t)=\rho_R\otimes_{j=1}^L\rho_{B_j}(t)\) where \(\rho_{B_j}(t)\) is the individual subsystem $j$, at certain time $t$ which we consider to be the equal mixture of low-lying state of the Hamiltonian $H_{B_j}$. In the  literature, \(\rho_{B_j}(t)=\ket{\psi_0}\bra{\psi_0}\) where \(\ket{\psi_0}\) is the ground state of \(H_{B_j}\). Let us describe the procedure to reach the final state in our case, which is \(\rho_{B_j}(t) = \frac{1}{k}\sum_{i=0}^{k-1} |i\rangle \langle i|\), i.e., we are interested to design a protocol which lead to the transformation, \(\rho(0) = \rho_R \otimes \rho_{in} \to \rho(t)=\rho_R\otimes_{j=1}^L(\frac{1}{k}\sum_{i=0}^{k-1} |i\rangle \langle i|)_j\).

	
	
	\begin{enumerate}
		\item {\it Unitary evolution.} After the preparation of the initial state, the system is evolved via an interacting Hamiltonian, \(H_{int}\) according to which all the \(B_{j}\)s interact and the regulator interacts with one of them or many of the $B_{j}$s. Therefore, the evolving Hamiltonian takes the form, as \(H_{tot}=H_{loc}+H_{int}\)  where $H_{loc}=\sum_{j}H_{B_j}$. Note that such interactions are accountable for creating correlations among all the qudits and the unitary operator \(U\) responsible for dynamics can be represented as \(U(t)=\exp(-iH_{tot}t)\).
		
		\item {\it Projective measurement with arbitrary rank.} After the evolution of the total system up to a finite time \(\tau\), a projective measurement of rank-$k$ is performed on the auxiliary system followed by a {\it postselection}. In order to take the inaccessible qudits to their low-energy sector, we choose the projectors to be formed by the eigenstate of the low-energy space of \(H_{B_j}\). Hence, rank-\(k\) projectors with \(1\leq k\leq d\) are given by 
		\begin{equation}
			P_1^{(k)}=\sum_{i=0}^{k-1}\ket{i}\bra{i},
		\end{equation}
		where the superscript $k$ denotes the rank of the projector while the subscript represents one of the element in  the measurement basis, \(\{P_1, P_2, \ldots\}\) where \(\sum_i P_i =\mathds{I} \) with \(\mathds{I}\) being the identity operator of the dimension of the auxiliary qudit and $\{\ket{i}\}_{i=0}^{d-1}$ are the eigenstates of the local Hamiltonian $H_{B_{j}}$. Accordingly, \(\rho_R\) is also prepared as \(\frac{1}{k}\sum_{i=0}^{k-1}\ket{i}\bra{i}\).
	\end{enumerate}
	The entire process, both unitary and rank-$k$ projective measurements, is repeated for a large number of time, N. Such procedure in literature is referred to as Zeno-like operation due to a finite non-vanishing time gap between two consecutive measurements. Mathematically, it can be written as
	\begin{equation}
		\rho(N\tau)=\frac{1}{p_1^{(k)}(N)}(P_1^{(k)}U(\tau))^N\rho(0)(U(\tau)^\dagger P_1^{(k)})^N,
		\label{eq:rho_t}
	\end{equation}
	where \(N\) is the number of measurements and \(p_1^{(k)}(N)\) is the probability to obtain \(P_1^{(k)}\), given by
	\begin{equation}
		p_1^{(k)}(N)=\Tr[(P_1^{(k)}U(\tau))^N\rho(0)(U(\tau)^\dagger P_1^{(k)})^N].
	\end{equation}
	To quantify how close the target state can be reached, we compute the Uhlmann's fidelity \cite{uhlmann1976,jozsa1994,mendon2008,miszczak2008} of each qudit, \(\rho_{B_j}\), after the $N$th measurement round and after tracing out all other qudits \(\{B_k\}_{k\ne j}\) and $R$.  Precisely, we calculate
	\begin{equation}
		\mathcal{F}\equiv\mathcal{F}_{B_j}^d(N,\{\alpha\},\tau)=\Big(\Tr\sqrt{\sqrt{\rho_{R}}\rho_{B_j}(N\tau)\sqrt{\rho_{R}}}\Big)^2,
		\label{eqn:U fidelity}
	\end{equation}
	where $\rho_{B_j(N\tau)}=\Tr_{\bar{B}}[\rho(N\tau)]$, where \(\bar{B}=\{B_k\}_{k\ne j}\) and, $\{\alpha\}$ denotes all the parameters involved in \(H_{tot}\) which can be tuned during dynamics and the initial state. Along with fidelity, it is also important to find out whether the probability of obtaining the target state after postselecting measurement is non-vanishing and finite, i.e., we are interested to compute \(p_1^{(k)}(N)\) which also guarantees the success of the protocol. In this work, we illustrate that there exists Hamiltonian and measurements operators, implementable in laboratories, that can lead to an almost unit fidelity with \(p_1^{(k)}(N)\ne0\).
	
	The entire procedure described above can also be interpreted from the perspective of local energy in the \(d\)-dimensional subsystem. Our goal is to lock the energy in the subspace, say, \((k\ll d))\). Hence, it is required to design a procedure such that the probability of projection of the subsystem upto the \(k^{\text{th}}\) energy subspace increases. To achieve this, one can apply the same procedure that we discuss above. Hence, instead of checking this energy-locking,
	we can compute the Uhlmann fidelity (\(\mathcal{F}\)) of the target systems obtained after \((P_1^{(k)} U(\tau))^N\)\cite{uhlmann1976,jozsa1994,mendon2008,miszczak2008} with \(\rho_R\) prepared initially as an equal mixture of the low-lying energy eigenstates.  When \(\mathcal{F}\to 1\), the successful locking  of energy or refrigeration occurs.

	Before moving towards the main results of our work, let us elaborate why such protocol works as presented in Refs.  \cite{nakazato2003, nakazato2004}. From Eq. (\ref{eq:rho_t}), we can observe that the repeated projection operation on the auxiliary system act as a nonunitary evolution on the inaccessible qudits which we wish to cool. Such nonunitary is represented as \(M_1^{(k)}=P_1^{(k)}U(\tau)\) and it acts on the inaccessible qudits \(N\) number of times. We can decompose \(M_1^{(k)}\) into spectral decomposition which is \(M_1^{(k)}=\sum_k \alpha_k \ket{R_k}\bra{L_k}\), where \(\ket{R_k}\) (\(\bra{L_k}\)) right (left) eigenvectors and \(\alpha_k\) are the  eigenvalues of \(M_1^{(k)}\) and \(\alpha_k\) are, in general, complex with \(0\le \abs{\alpha_k} \le 1\). Now successive applications of \(M_1^{(k)}\), the system approaches to the eigenstate corresponding to the right eigenstate of \(\alpha_k^{\max}\) as  \(N\to\infty\) while  the other \(\abs{\alpha_k^{\ne \max}}\to 0\). Note that \(\alpha_k^{\max}\) as well as \(\ket{R_k^{\max}}\) (\(\bra{L_k^{\max}}\)) depend upon the different system parameters and measurement operators, which may make \(\ket{R_k^{\max}}\) entangled or product state. Hence, we need to find out accurate projectors as well as system parameters such that system can act as refrigerator which we will analyze explicitly in our scenario in succeeding sections.

	\section{Subspace cooling with paradigmatic spin models}
	\label{sec:paradigmatic}

	To illustrate the benefit of higher-dimensional system for building quantum refrigerator, we consider two paradigmatic spin-\(s\) models, namely \(XXZ\) and bilinear biquadratic Heisenberg $(BBH)$ spin models, for obtaining the dynamical state. Moreover, the role of applying rank-\(k\) projectors repeatedly on the regulator  for cooling of the individual qudit into low-lying states will also be addressed here. First, we show that independent of the rank of the projectors, each qudit can be cooled.
	
	
	
	Let us demonstrate the traits of refrigerator where the target system is made of a single qudit, i.e., \(L=1\). It is initially prepared in a maximally mixed state, i.e., the thermal state with $\beta=0$.  On the other hand, the auxiliary qudit is initially either in the ground state of the local Hamiltonian, i.e., \(\rho_R=\ket{0}\bra{0}\) or  \(\rho_R=\frac{1}{k}\sum_i\ket{i}\bra{i}\) (\(k \ll d\))  in which the target qudit has  to be transformed at the end of the protocol. Therefore, the total system is given as 
	\begin{equation}
		\rho(0)=\rho_{R}\otimes\frac{\exp(-\beta S_1^z)}{\Tr[\exp(-\beta S_1^z)]}\Big|_{\beta=0}\equiv\rho_R\otimes\rho_{B_1}^{\beta=0},
	\end{equation}
	where \(S_1^z\) is the local Hamiltonian of the inaccessible target qudit which is spin-\(s\) operators in \(d\)-dimension in \(z\)-direction. After preparation of the initial state, both the qudits undergo an evolution governed by the Hamiltonian,
	\begin{eqnarray}
		\nonumber H_{XXZ}^d &=&\sum_{j=1}^{L}J'\Big[S_{j}^x S_{j+1}^x+S_{j}^y S_{j+1}^y+\Delta S_{j}^z S_{j+1}^z\Big]
		\nonumber\\&+& h'\sum_{j=1}^{L+1}S_{j}^z
		\label{eq:hamil spin-d Heisen}
	\end{eqnarray}
	where \(S^{i}\) (\(i\in \{x,y,z\}\)) are the spin-\(s\) operators in \(d\)-dimension, \(h'\) is the strength of the local magnetic field, \(J'\) is the interaction strength in \(xy\)-plane between two nearest-neighbor qudits, and \(\Delta\) is the interaction strength towards the \(z\)-direction. The Hamiltonian is known as the $XXZ$ model in $d$ dimension, which can possess exotic properties in \(d>2\) \cite{dutta}. Note that the regulator qudit is initialized as the ground state or low-lying state of the local part of \(H_{XXZ}^d\) (the second term in Eq. (\ref{eq:hamil spin-d Heisen})). 
	
	Another  spin-\(s\) Hamiltonian, given as
	\begin{eqnarray}
		\nonumber H_{BBH}^d&=&J'\sum_{j=1}^{L}\cos\theta \vec{S_j}.\vec{S}_{j+1}+\sin\theta(\vec{S_j}.\vec{S}_{j+1})^2
		\\&+& h'\sum_{j=1}^{L+1}S_{j}^z
		\label{eq:hamil_bbh}
	\end{eqnarray}
	is also used to evolve the regulator and the target qudit. The Hamiltonian in Eq. (\ref{eq:hamil_bbh}) is referred to as  bilinear and biquadratic Heisenberg Hamiltonian, in short \(BBH\) model which has no counterpart in spin-\(1/2\) system. Here $\theta$ is the system parameter, we name this as {\it phase parameter}, which basically governs the ratio between bilinear and biquadratic terms. This nearest-neighbor one-dimensional (1D) \(BBH\) model shows a rich phase structure with various disordered phase with respect to $\theta$. Specifically, three different phases, namely the trimerized , the Haldane  and the dimerized phases emerge.  And the only order phase appears in this model is ferromagnetic, as its order parameter $S_z$ is conserved. So the four quantum phases in this model are the ferromagnetic phase $(-\frac{3\pi}{4}\leq\theta\leq -\pi,\frac{\pi}{2}\leq\theta\leq\pi)$, the critical (trimerized) phase $(\frac{\pi}{4}\leq\theta\leq\frac{\pi}{2})$, the Haldane phase $(-\frac{\pi}{4}\leq\theta\leq\frac{\pi}{4})$, and the dimerized phase $(-\frac{3\pi}{4}\leq\theta\leq-\frac{\pi}{4})$\cite{BBH2022}. 
	We will exhibit that the subspace cooling depends on the phase parameters chosen to evolve the system.

	\subsection{Two-qudit refrigerator using rank-1 measurement}
	\label{subsubsec:rank2}
	
	Let us first describe the situation where we perform only rank-\(1\) measurement and hence the target state at \(B_1\) gets purified to a pure state. In particular, we are interested in the transformation, \(\rho(0) \underset{(P_1^{(1)}U)^N}{\Rightarrow} \rho(N\tau)\) when the outcome of the measurement is \(|0\rangle \langle 0|\). For evolution, we choose spin-\(s\) \(XX\) model, i.e.,  \(H_{XXZ}^d\) having \(\Delta=0\) and \(BBH\) model with different \(\theta\) parameters. 
	
\begin{figure}
		\centering
		\includegraphics[scale=0.38]{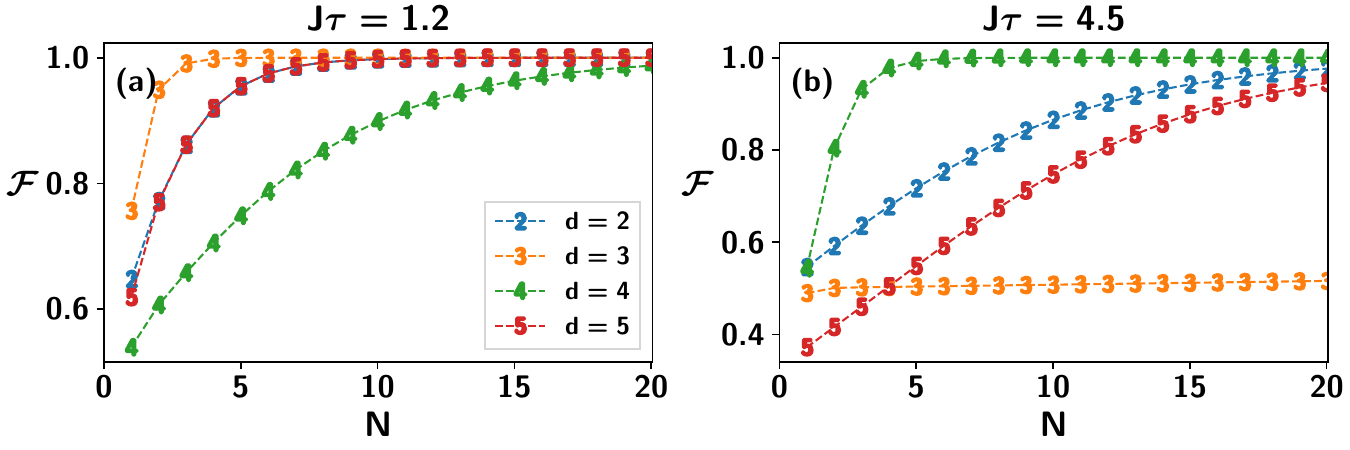}
		\caption{(Color online.) The fidelity of the qudit $B_1$, $\mathcal{F}$ (vertical axis) against the number of rank-$1$ projective measurement along with unitary evolution, $N$ (horizontal axis) for the $XX$ Hamiltonian. Different color signifies different  dimension $d$ of  $B_1$. In (a), $J\tau =1.2$ and in (b) $J\tau =4.5$ of the \(H_{XXZ}^d\) with \(\Delta =0\). Note that depending on the values of $J\tau$, $\mathcal{F}$ saturates to unity for a fixed dimension $d$. All the axes are dimensionless.}
		\label{fig:ldim advantage_2}
	\end{figure}

	{\it Refrigeration with \(XX\) model.} After the system evolves for a time period with \(H_{XXZ}^d\) having \(\Delta=0\), the rank-\(1\) measurement is performed in the computational basis in the d-dimension, i.e., \(\{\ket{i}_{i=0}^{d-1}\}\) is carried out on the regulator qudit. As described in Sec. \ref{sec:framework}, the evolution is performed for \(\tau\) time period followed by a measurement  and the entire process is performed repeatedly for \(N\) times. To assess the performance of the procedure, we determine the fidelity between the desired state, \(\ket{0}\) and the target qudit after \(N\) steps, \(F_{B_1}^d=\bra{0}\rho_{B_1}(N\tau)\ket{0}\) for a fixed dimension \(d\). In case of spin-\(1\) system (i.e., for \(d=3\)), the fidelity expression reads as
	\begin{equation}
		\mathcal{F}_{B_1}^{3}(N,J\tau) = \Big(1+\cos^{2N}J\tau+\cos^{4N}\frac{J\tau}{\sqrt{2}}\Big)^{-1},
		\label{eq:fid_d_1}
	\end{equation}
	where \(J=J'/h'\). Since \(\cos^2J\tau\le 1\), we note that \(\cos^{2N}J\tau\to 0\) and \(\cos^{4N}\frac{J\tau}{\sqrt{2}}\to 0\) as \(N\to\infty\) provided \(J\tau\) is chosen appropriately and in this situation, \(\mathcal{F}_{B_1}^{3}\) approaches to unity. Further, we notice that for \(J\tau=\pi\), the fidelity \({F}_{B_1}^{3}\to 0.5\) for \(N\to\infty\). Hence, for certain values of the interaction strength along with time period \(\tau\), the fidelity never approaches to unity even with high \(N\) (see Fig. \ref{fig:ldim advantage_2}). However, practical implementation demands \(N\) to be finite. Moreover, we observe that \(\mathcal{F}_{B_1}^{3}\) is periodic function of \(J\tau\). It is worthwhile to mention here that the expression for fidelity involves a higher number of trigonometric terms with the increase of dimension. E.g. \(d=2\), \({F}_{B_1}^{2}\) has the form \((1+\cos^{2N}2J\tau)^{-1}\). A systematic analysis reveals that
	\begin{widetext}
		\begin{eqnarray}
			\label{eqn:F4 fidelity}\mathcal{F}_{B_2}^4(N,J\tau)& = &\Big(1+\cos(\frac{3J\tau}{2})^{2N}+\cos(\sqrt{\frac{3}{2}}J\tau)^{4N}+\Big[\cos(J\tau)\cos(\frac{\sqrt{13}J\tau}{2})+\frac{2}{\sqrt{13}}\sin(J\tau)\sin(\frac{\sqrt{13}J\tau}{2})\Big]^{2N}\Big)^{-1},\\ \label{eqn:F5 fidelity} \, \text{and}\,  \mathcal{F}_{B_2}^5(N,J\tau)&=&\Bigg(1+\cos(2J\tau)^{2N}+\cos(\sqrt{3}J\tau)^{4N}+\frac{1}{22^{2N}}\Big(9+11\cos(2J\tau)+2\cos(\sqrt{22}J\tau)\Big)^{2N}\\
			\nonumber &+&
			\frac{1}{22^{2N}}\Big((11+\sqrt{33})\cos(\frac{1}{2}(3-\sqrt{33})J\tau)+(11-\sqrt{33})\cos(\frac{1}{2}(3+\sqrt{33})J\tau)\Big)^{2N}\Bigg )^{-1}.
		\end{eqnarray}
	\end{widetext}
	
	From the above expressions, it is clear that \({F}_{B_2}^{2}\) is, in general, periodic with variation of \(J\tau\) and \(N\) although the periodicity become nontrivial with the increase of dimension. The trends of, \({F}_{B_2}^{d}\) by varying \(J\tau\) and \(N\), reveal that there exists a range of \(J\tau\) values for which \({F}_{B_2}^{d}\) never achieves unity even with the increase of \(N\) as depicted in Fig. \ref{fig:ldim advantage_2}, irrespective of the dimension. We refer to such a range of \(J\tau\) values as imperfect refrigerating parameters. Interestingly and unfortunately, the range of imperfect  refrigerating parameters get increased with the increase of dimensions. The above study possibly indicates that with the increase of dimension, purification to a ground state is not easy.  We will exhibit in the next subsection that the dimensional gain can be achieved when one relaxes the cooling procedure from the ground state to the low-lying states.

	{\it Refrigerator with bilinear-biquadratic interactions. } 
	
\begin{figure}
		\centering
		\includegraphics[scale=0.38]{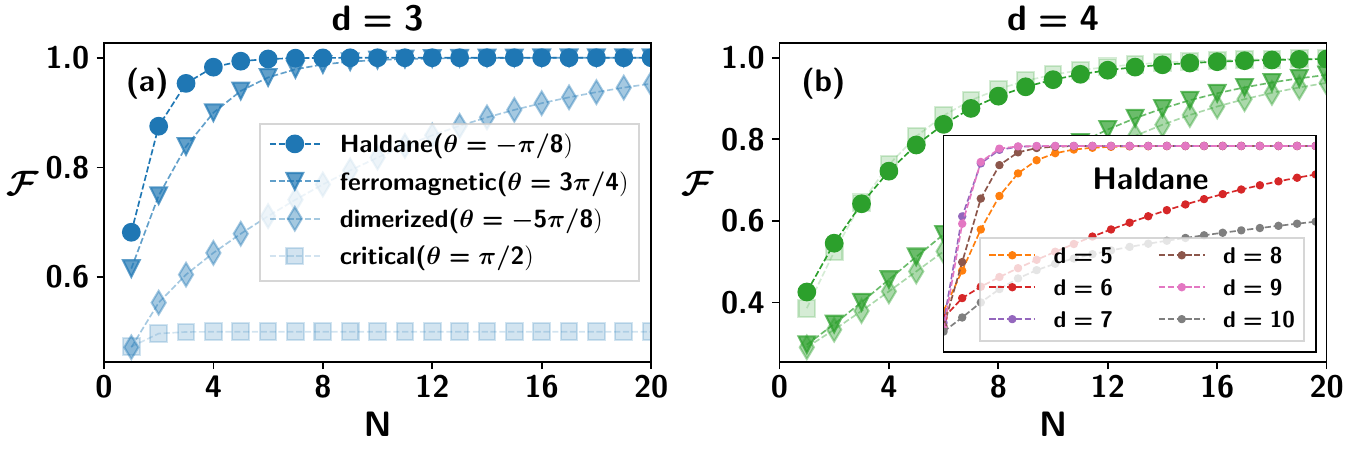}
		\caption{(Color online.)  $\mathcal{F}$ (vertical axis) against  $N$ (horizontal axis) for the $BBH$ Hamiltonian  having different values of $\theta$  with  $J\tau=1$. Again the rank-\(1\) measurement is performed. Different shades and shapes of blue and green represent $d=3$ and  $d=4$ respectively. Note that different  \(\theta\) values signify  different phases of the \(BBH\) model. (a) For $d=3$, fixing $\theta=\frac{\pi}{2}$ (critical phase), $\mathcal{F}$ never reaches to unity while  $\theta=-\frac{\pi}{8}, \frac{3\pi}{4}, \frac{-5\pi}{8}$ signifying Haldane, ferromagnetic and dimerized  phases respectively, unit fidelity is achieved. Interestingly, when $d=4$, the unit fidelity is attained irrespective of \(\theta\) values although in the ferromagnetic and Haldane phases, \(N\) is smaller than the one  obtained with \(\theta\) belonging to the dimerized and critical phases.  All the axes are dimensionless.}
		\label{fig:BBQ_phase_2}
	\end{figure}

	Using the Zeno-like approach, we also investigate the performance of the two-spin refrigerator when the spin-spin interactions are governed by the \(BBH\) Hamiltonian. Such a study also helps us establish that the protocol is precisely independent of the choice of the evolving Hamiltonian. Choosing rank-\(1\) measurement on the auxiliary system, we end up to calculate the fidelity of the target state, given by
	\begin{eqnarray}
		\nonumber\mathcal{F}_{B_2}^3(N,\theta,J\tau)&=&\Big(1+\cos(a_{10}J\tau)^{2N}\\\nonumber&+&\frac{1}{18^N}\Big[7+3\cos(2a_{10}J\tau)\\&+&6\cos (a_{13}^-J\tau)+2\cos(3a_{11}^{-}J\tau)\Big]^N\Big)^{-1},\nonumber\\
	\end{eqnarray}
	where \(a_{mn}^{\pm}=m\cos\theta\pm n\sin\theta\)  and \(a_{m0}^{\pm}=a_{m0}=m\cos\theta\) and \(\theta\) is the parameter dictating the phases of the system, $J=\frac{J'}{h'}$. From the expression of fidelity, it can be observed that for particular choice of \(J\tau\) and \(\theta\), the unit fidelity can be achieved, although depending on the value of $\theta$, $\mathcal{F}_{B_1}^3$ scales differently. For example, we observe that the fidelity behaves differently  in the ferromagnetic phase (as shown in Fig. \ref{fig:BBQ_phase_2} with $\theta=\frac{3\pi}{4}$) than the one when the evolving Hamiltonian is in the Haldane phase   (see Fig. \ref{fig:BBQ_phase_2} with  $\theta=-\frac{\pi}{8}$) with respect to \(N\) required to achieve unit fidelity. Moreover, the fidelity changes its characteristics drastically when \(d\) grows.

	\subsection{Refrigeration with rank-$2$ projector}
	
\begin{figure*}
		\centering
		\includegraphics[width=\textwidth]{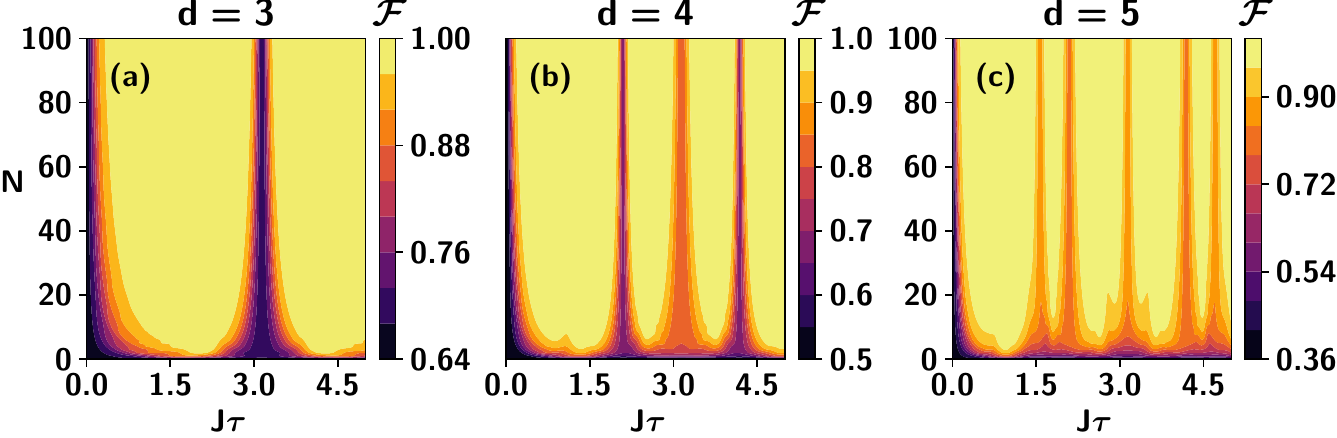}
		\caption{(color online.) Contour plot of fidelity, \(\mathcal{F}\) of qudit $B_{1}$ with respect to the interaction strength times time period, \(J\tau\) (abscissa) and  $N$ (ordinate) for different Hilbert space dimension of the individual target spin, $d$. Figs. (a), (b), and (c) indicate \(d= 2, 3,\) and \(4\) respectively. The associate interacting  Hamiltonian used for evolution is the $H_{XXZ}$ model with $\Delta=1$. Unlike Figs. \ref{fig:ldim advantage_2} and \ref{fig:BBQ_phase_2},   rank-$2$ projection operators \(P_1^{(2)} = |0\rangle \langle 0| + |1\rangle \langle 1| \) is used and the initial regular state  is taken as \(\frac{1}{2}(|0\rangle \langle 0| + |1\rangle \langle 1|) \). Note than there are certain regions  of \(J\tau, N\)-plane in every sub-figures where $\mathcal{F}>0.96$ which indicates that quantum refrigerator can work efficiently. It is interesting to notice that for $J\tau\approx 3.0$ in $d=3$ or $J\tau\approx 2.0$ in $d=4,$ and \(5\),  it is impossible to achieve perfect cooling (unit fidelity or even high fidelity) irrespective of the values of $N$. All the axes are dimensionless.}
		\label{fig:fidelity_N_J}
	\end{figure*}

\begin{figure}
		\centering
		\includegraphics[scale=0.5]{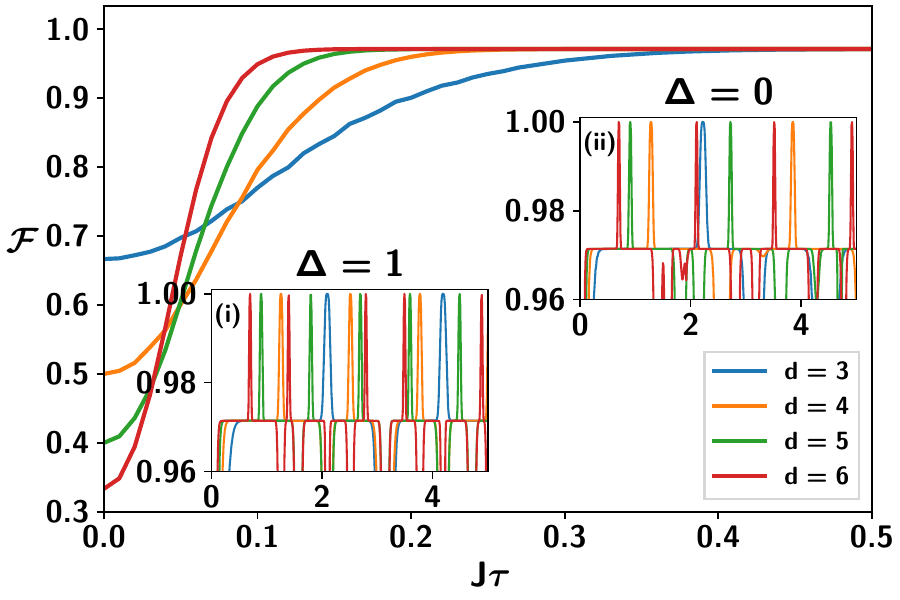}
		\caption{(Color online.)  $\mathcal{F}$ (vertical axis) vs \(J\tau\) (horizontal axis) for different values of \(d\). 
			The evolution is done by the  $H_{XXZ}^d$ Hamiltonian with $\Delta=1$ and the fidelity is computed after $N=100$  with rank-$2$ projectors mentioned in Fig. \ref{fig:fidelity_N_J}. Starting from an initial value,  $\mathcal{F}$ reaches to unity depending on $d$. Both the inset (a) for $\Delta=0$ and (b) for $\Delta=1$ show that $\mathcal{F}$ hits unity after $J\tau>0.5$. All the axes are dimensionless.}
		\label{fig:fidelity_J}
	\end{figure}

	Instead of demanding the unmeasured qudits in the ground state by performing repeated local measurements on the regulator, we opt for a less stringent transformation. Specifically, we wish to activate transition: \(\rho(0)=\rho_R\otimes_{j=1}^L\rho_j\to\rho(N\tau)=\rho_R\otimes_{j=1}^L\rho_R\) where \(\rho_R=\frac{1}{k}\sum_{i=0}^{k-1} \ketbra{i}{i}\) with \(k\ll d\). Let us illustrate the situation when  $k=2$ in which the measurement is performed in the basis, \(\{P_1^{(2)}=\ket{0}\bra{0}+\ket{1}\bra{1}, P_2,P_3,\dots\}\) and we are interested to the probability and fidelity of obtaining \(P_1^{(2)}\). Note that such higher rank projectors can only be used in \(d>2\) and hence the procedure described for subspace cooling or locking of states in low-lying energy sector can only work in higher dimension. Precisely, we   prepare the regulator qudit with an equal mixture of the ground and the excited state of the local Hamiltonian \(S_j^z\), i.e., \(\rho_R=\frac{1}{2}(\ket{0}\bra{0}+\ket{1}\bra{1})\) while the target qudit is in a maximally mixed state which was the same as considered in the previous case.
	In order to refrigerate the inaccessible qudit following the Zeno-like protocol, we evolve the system either with \(H_{XXZ}^d\) or with \(H_{BBH}\) models. In both the situations, we are interested with the outcome of the measurement being \(P_1^{(2)}=\ket{0}\bra{0}+\ket{1}\bra{1}\). In the case when the evolution is governed by the $XXZ$ model, the fidelity expression of the individual target state, after $N$ number of rounds, takes the form 
	\begin{equation}
		\mathcal{F}_{B_1}^d(N,J\tau)=\frac{\Big(C(d,N)+\sqrt{\sum_{r=0}^{3N}A_r(d,N)\cos(rJ\tau)}\Big)^2}{\sum_{r=0}^{3N}A_r^{'}(d,N)\cos(rJ\tau)},
		\label{eqn:fidelity subspace}
	\end{equation}
	where \(C(d,N)\), \(A_r(d,N)\) and \(A'_r(d,N)\) are constants for fixed $d$ and $N$. We skip the exact forms of these constants since depending on $d$ and $N$, they change and become cumbersome. For a fixed $N$, we observe that $\mathcal{F}_{B_1}^d$ approaches unity for all values of $d$ (see Fig. \ref{fig:fidelity_N_J} by varying both \(J\tau\) as well as \(N\) and Fig.  \ref{fig:fidelity_J}   with a fixed \(N=100\) with respect to  $J\tau$)). Interestingly, we find that for a fixed $N$, the rate of increase in \(\mathcal{F}_{B_1}^d\) with \(J\tau\) increases  with the increase of $d$, although all of them saturates to a constant value which is independent of the dimension. As seen before with rank-\(1\) measurement, the behavior of the fidelity  depends highly on the evolving Hamiltonian. For example,  the expression of fidelity for spin-\(1\) system reads as
	\begin{align}
		\nonumber &\mathcal{F}_{B_1}^{3}(N,J\tau)=\Big[2^{(N-\frac{1}{2})}\Big(\sum_{r=0}^{N}A_{r}^1(N)\cos(2rJ\tau)\\\nonumber&+\sum_{r=1}^{2N}A_{r}^2(N)\cos(\sqrt{2}rJ\tau)\Big)^{-\frac{1}{2}}+\\&\Big(\frac{\sum_{r=0}^{N}A^3_r(N)\cos(2rJ\tau)+\sum_{r=1}^{2N}A^4_r(N)\cos(\sqrt{2}rJ\tau)}{\sum_{r=1}^{2N}A^5_r(N)\cos(\sqrt{2}rJ\tau)}\Big)^{-\frac{1}{2}}\Big]^2,\nonumber\\
		\label{eqn:fidelity3 subspace} 
	\end{align} 
	where $A_r^i (i=1,\ldots,5)$ are the constants which depends on the parameters of the evolving Hamiltonian, \(XXZ\) model (see the inset of Fig. \ref{fig:fidelity_J} for $\mathcal{F}_{B_1}^d$ when $XXZ$ Hamiltonian with $\Delta=0$ is used for evolution.)
	
	{\it Rank-$2$ measurement and $BBH$ Hamiltonian for evolution}. Let us concentrate when the $BBH$ model  is applied to evolve the system.  In particular, the initial state and the measurement remain the same as before, only changes happen in the evolving Hamiltonian. As shown in Fig. \ref{fig:BBQ_phase_2} in the case of rank-$1$ measurement, the fidelity depends on the choice of the parameter $\theta$ in $H_{BBH}$. Note that the  expression fidelity, in this case becomes more cumbersome than the one obtained for rank-\(1\) measurement. Precisely, we find that for $d=3$ and $N=2$, the fidelity reads
	\begin{eqnarray}
		\nonumber &&F_{B_1}^3(2,J\tau)=\\\nonumber&&\Big[\Big(\frac{\sum_{m,n,\pm}C_1(\pm,m,n)\cos(C_1^{'}(\pm,m,n)a_{mn}^{\pm}J\tau)}{\sum_{m,n,\pm}C_2(\pm,m,n)\cos(C_2^{'}(\pm,m,n)a_{mn}^{\pm}J\tau)}\Big)^{-\frac{1}{2}}\\&+&\Big(\sum_{m,n,\pm}C_3(\pm,m,n)\cos(C_3^{'}(\pm,m,n)a_{mn}^{\pm}J\tau)\Big)^{-\frac{1}{2}}\Big]^2,\nonumber\\
	\end{eqnarray}
	where \(C_i(\pm,m,n)\) and \(C^{'}_i(\pm,m,n)\) are the constants  which depends on the system parameters.

	\subsection{Probability dependency on spin dimension}
	
\begin{figure}
		\centering
		\includegraphics[scale=0.5]{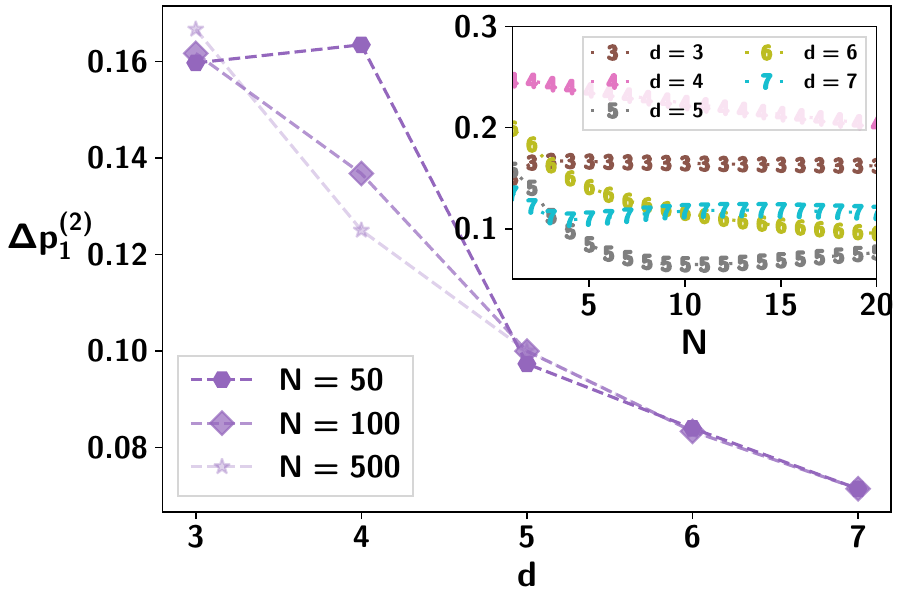}
		\caption{(Color online.) Difference between success probability obtained with rank-\((k-1)\) and rank-\(k\) projectors, $\Delta p_1^{(k)}=p_1^{(k)}-p_1^{(k-1)}$ (vertical axis) with respect to the dimension of the individual target system, $d$ (horizontal axis). Here $k=2$ and  different $N$ are chosen for illustration. Interestingly, $\Delta p_1^{(2)}$ shows nontrivial behavior, when it increases for $N=50$ at $d=4$ instead of decreasing. In the inset, the same quantity, $\Delta p_1^{(2)}$ (\(y\)-axis) against $N$ for different $d$  values. It clearly displays a nontrivial behavior of $\Delta p_1^{(2)}$ as it scales differently with $N$ depending on $d$. Notice that different symbols  indicate different $N$  while different symbol in the inset is for different dimension. All the axes are dimensionless.}
		\label{fig:adv prob}
	\end{figure}

	One of the important quantity, which also quantifies the effectiveness of the refrigeration process, is the probability to obtain the postselected projector $P_1^{(k)}$. The probability of obtaining the outcome $P_1^{(k)}$ depends on the dimension for fixed $J\tau$ and $N$. In the picture of subspace cooling, one can expect that the probability should increase with the rank of the projector $P_1^{(k)}$. Further, when $d$ and $k$ are comparable, the advantage of rank-$k$ projectors over rank-$1$ projectors of $P_1^{(k)}$ in terms of probability and fidelity can be prominent (see Fig. \ref{fig:adv prob} and Fig. \ref{fig:rank_k_prob}). To make the analysis more quantitative, we define a quantity which is a difference between the probability of obtaining $P_1^{(k)}$ with rank-$k$ and rank-$(k-1)$ as $\Delta p_1^{(k)}=p_1^{(k)}-p_1^{(k-1)}$. One expects that it should decrease with \(N\) and with the increment of dimension. It indeed decreases with \(d\) as shown in Fig. \ref{fig:adv prob} which depicts the behavior of $\Delta p_1^{(2)}$ by varying $d$ for the $XXZ$ model as the evolving Hamiltonian. This is because the probability of obtaining  $P_1^{(2)}$ and  $P_1^{(1)}$  becomes almost same when \(d\) increases although they may provide nontrivial inference when \(d\) is small. On the other hand, this difference in probability shows nonmonotonic behavior with the increase of \(N\), especially when \(N\) is not too large (see the inset of  Fig. \ref{fig:adv prob}).

	\subsection{Refrigeration with rank-\(k\) projectors}
	\label{subsubsec:rank k}
	
\begin{figure}
		\centering
		\includegraphics[scale=0.38]{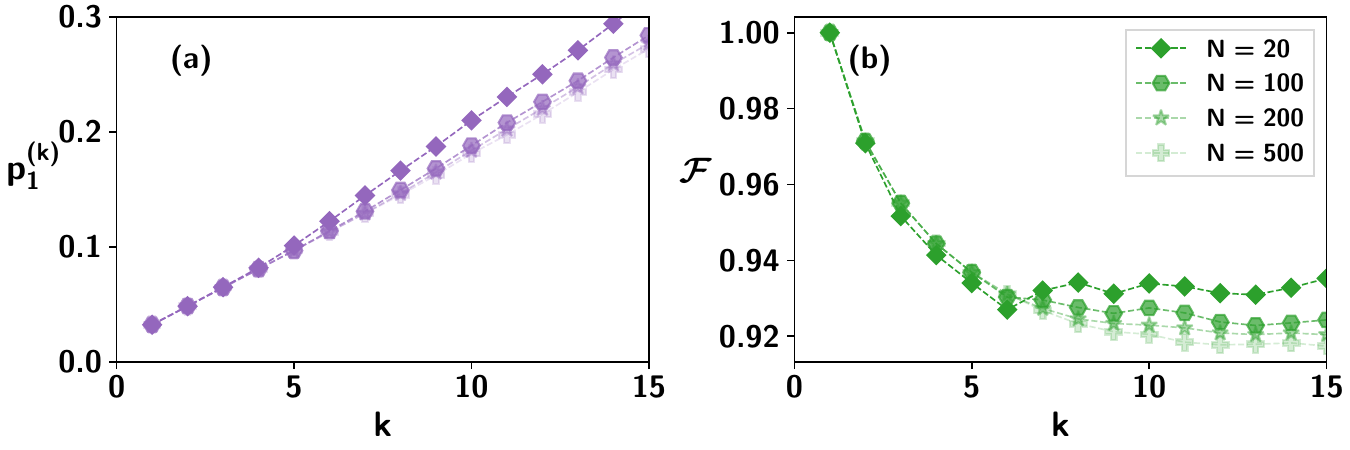}
		\caption{(Color online.) (a) The success probability of the refrigeration protocol, $p_1^{(k)}$ (vertical axis) against the rank of the projective measurement, $k$ (horizontal axis). Different shapes and shades  indicate different  $N$. $p_1^{(k)}$ increases almost linearly with rank, $k$, of the projectors where different slopes are for different \(N\). Here  $d=31$ where $k < (d+1)/2$. The evolving Hamiltonian is chosen to be the \(XXZ\) model and \(J\tau=3\).  (b) Fidelity, $\mathcal{F}$ (\(y\)-axis) with respect to the rank of the projective measurement, $k$ (\(x\)-axis). It behaves oppositely in contrast to that $p_1^{(k)}$ for different $N$ values. 
			Comparing (a) and (b), we also notice an interplay between the fidelity of the target system with the success probability.   Dark to light shades of color indicates the increase of  $N$. All the axes are dimensionless.}
		\label{fig:rank_k_prob}
	\end{figure}

	To probe the role of rank in the measurement for a fixed dimensional refrigerator, one requires to increase the dimension of the individual system. Towards that aim, let us choose the dimension \(d\) to be large, and we can vary the rank of the projectors, \(P_1^{(k)}\) up to \(d/2\) (\(d+1/2\)) when \(d\) is even (odd). 
	When \(k>d/2\), the fidelity again starts to increase since the initial state of the inaccessible qudit is initially taken to be maximally mixed state and with the increment of the rank of the projector, the regulator also becomes close to \(\mathbb{I}/d\), thereby attaining to the unit fidelity for a fixed \(N\). Therefore, our aim is to successfully perform the transformation,
	\begin{equation}
		\frac{1}{k}\sum_{i=0}^{k-1} \ketbra{i}{i} \otimes \frac{\mathbb{I}}{d}\underset{(P_1^{(k)} U)^N}{\Rightarrow}\frac{1}{k}\sum_{i=0}^{k-1} \ketbra{i}{i} \otimes \frac{1}{k}\sum_{i=0}^{k-1} \ketbra{i}{i}.
	\end{equation}
	Let us enumerate the observations in this situation.
	\begin{enumerate}
		\item For a fixed \(d\), and when the evolution Hamiltonian is fixed, the trends of the probability \(p_1^{(k)}\) and the fidelity does not depend on \(N\), provided \(N\) is moderately high.
		
		\item The probability of obtaining the outcome \(P_1^{(k)}\) increases with \(k\) as illustrated in Fig. \ref{fig:rank_k_prob}(a) when the evolution Hamiltonian is taken to be \(XXZ\) with \(d=31\). However, the behavior does not alter drastically for other evolving spin model.
		
		\item The pattern of fidelity with \(k\) is opposite than the one observed for \(p_1^{(k)}\), i.e., \(F_{B_2}^d\) decreases with \(k\). Interestingly, \(\exists\) a critical \(k\) after which the fidelity remains constant for moderately high values of \(N\). For example, \(d=31\), we find that \(k\geq 7\), the fidelity remains almost unaltered.   
		
	\end{enumerate}
	These results can be summarized in the following observation: \\
	\textbf{Observation.} For a given dimension \(d\) of the individual inaccessible target system, the subspace cooling occurs only when the interaction strength, the time period up to which the regulator and the target systems interact, and the number of repeated measurements are chosen suitably. All these parameters may not remain appropriate when the dimension of the individual site varies (see Fig. \ref{fig:fidelity_N_J}). On the other hand, the probability of obtaining the higher rank projector gets enhanced with the increase of the rank of the measurement (see Fig. \ref{fig:rank_k_prob}).

	\section{Refrigeration made of \(L\)-array of qudits.}
	\label{sec:l_array}
	
	Until now, we discuss a quantum refrigerator made up of two subsystems, a target system to be refrigerated, and other qudit is used as auxiliary system which is used to cool the target. Let us now increase the number of target subsystems, i.e., we are interested in refrigeration of \(L\)-array of qudits where each qudit individually is in a thermal state with inverse temperature \(\beta =0\). In this scenario, a regulator qudit is attached in the boundary and the refrigeration protocol is to bring the individual qudit in their respective low energy space via interacting Hamiltonian. Precisely, our aim is to transform qudits in the following manner:
	\begin{equation}
		\frac{1}{k}\sum_{i=0}^{k-1} \ketbra{i}{i} \bigotimes_{j=1}^{L}\frac{\mathbb{I}_j}{d}\underset{(P_1^{(k)}U)^N}{\Rightarrow}\frac{1}{k}\sum_{i=0}^{k-1} \ketbra{i}{i} \bigotimes_{j=1}^{L} \Big [\frac{1}{k}\sum_{i=0}^{k-1} \ketbra{i}{i}\Big ]_j.
	\end{equation}

	\subsection{Refrigeration using rank-2 measurement}
	\label{subsec:rank2 array}
	
	The main motivation is to investigate how the refrigeration process gets affected when one increases the number of target subsystems. Let us fix the dimension, \(d=3\) and the rank of the projector, \(P_1^{(2)}\) to be two. When the entire system evolves according to \(H_{XXZ}\) model in Eq. (\ref{eq:hamil spin-d Heisen}) and \(H_{BBH}\) in Eq. (\ref{eq:hamil_bbh}) for $d=3$, the fidelity exhibits some appealing behavior. Suppose, we assume that the first target system is in contact with the regulator  and the farthest qudit is the \(L\)-th one. We first notice that the fidelity of the first target qudit reaches its maximum value while the farthest qudit has the minimum fidelity with desired state. Secondly, we observe that with \(L=4\), the fidelity of the farthest qudit never achieves unity for any value of \(N\), \(J\tau\) and \(\theta\) (see Fig. \ref{fig:BBQ_phase_2}) in which we fix \(F_{B_4}^{3}\le 0.9\). The situation deteriorates with the increase of \(L\). 
	
	\begin{figure}
		\centering
		\includegraphics[scale=0.38]{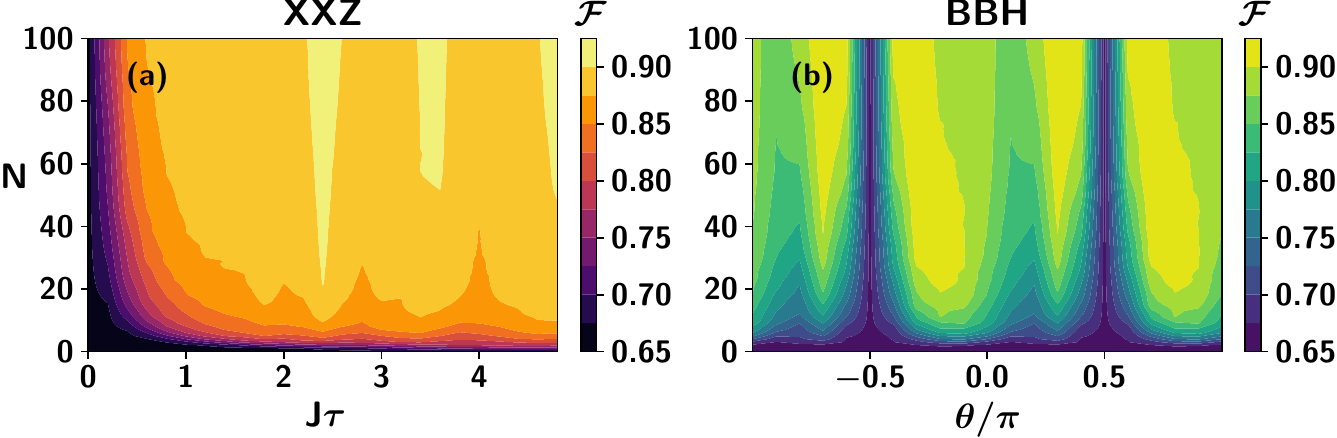}
		\caption{(color online.) (a) Contour plot of fidelity, \(\mathcal{F}\) of the farthest qudit, $B_4$ from the regulator  with respect to $N$ (\(y\)-axis) and  \(J\tau\) (\(x\)-axis) for spin-$1$ $XXZ$ interacting Hamiltonian  with $\Delta=1$. Here $L=4$ and the rank of the projector is  $k=2$.  (b) The same for the \(BBH\) Hamiltonian chosen for evolution and the horizontal  axis represents the phase parameter, \(\frac{\theta}{\pi}\) (\(x\)-axis). We choose \(J\tau=1\).  All the axes are dimensionless.}
		\label{fig:XXZ_BBG}
	\end{figure}
	
	
	
	\section{Refrigeration in Star configuration}
	\label{sec:lattice geometry}
	
	We discuss the effect of geometry in the refrigeration protocol. In particular,  we take a system of a spin-star model \cite{Mahesh2021_startopology} which is a central spin model of local spin dimension \(d\). In this model, \(L\) number of qudits interact with a single central spin of dimension \(d\) (see Fig. \ref{fig:schematic} (b)). For spin-\(1/2\) system, such model is realized in an experimental setup to achieve refrigeration in experiment \cite{mahesh_star_cool}. The model reads as
	\begin{equation}
		H_{SS}=h'S_R^z +J'\sum_{i=1}^{L} (S_R^xS_i^x+S_R^yS_i^y).
	\end{equation}
	Here \(S_R^{i}\), \(i\in{x,y,z}\) is the local spin Hamiltonian of the central-spin and \(h'\) is the local magnetic field for the central spin which is given as \(J=J'/h'\) where \(J\) is the interaction strength between the central spin and spins in the ring. In the spin-star model,  we perform rank-\(k\) measurement on the central spin which, in this case, acts as a regulator and the central spin is initially prepared as \(\frac{1}{k}\sum_{i=0}^{k-1} |i\rangle \langle i|\) and the rest of the qudits in the ring  are in the thermal state with infinite temperature.
	
\begin{figure}
		\centering
		\includegraphics[scale=0.38]{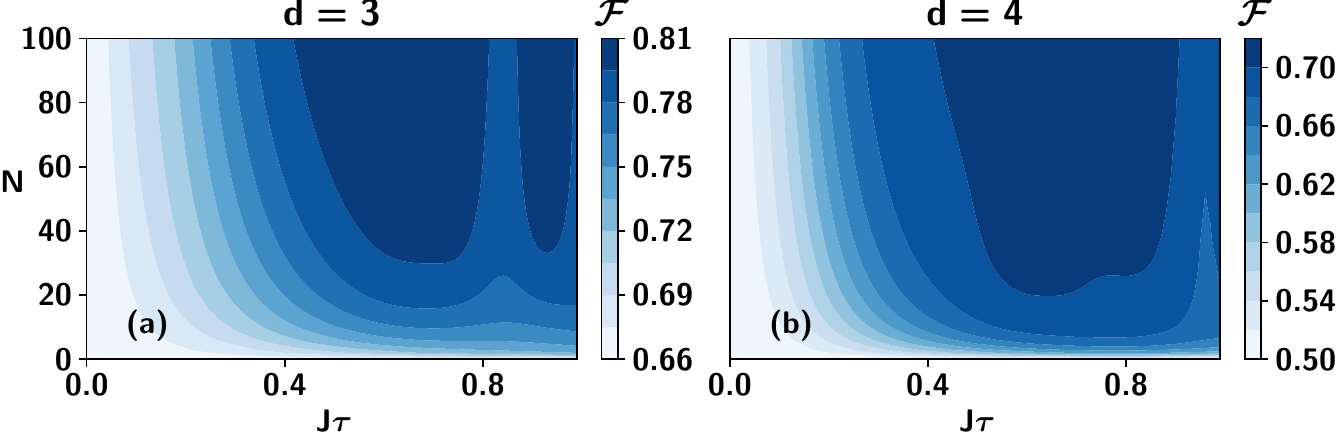}
		\caption{Contour plot of fidelity, \(\mathcal{F}\),  is plotted against  \(J\tau\) (ordinate) and \(N\) (abscissa). Here \(L=4\) and the evolution Hamiltonian is chosen to be the \(XX\) model, i.e., \(H_{XXZ}^d\) with \(\Delta =0\). The regulator qudit is the central spin which is surrounded by four qudits and the rank-2 projector, \(P_1^{(2)}\) mentioned in Fig. \ref{fig:fidelity_N_J} is obtained. Interestingly, we notice that unlike all the previous cases, the fidelity never reaches to unity irrespective of the dimension, \(J\tau\) and \(N\) due to the symmetry present in this system.   All axes are dimensionless.}
		\label{fig:enter-label}
	\end{figure}

	Let the evolution and  measurements be performed \(N\) times, given as \((P_1^{(k)}U(\tau))^N\).  We  calculate the fidelity of each qudit in the ring.  Unlike previous cases discussed (in Sec. \ref{sec:l_array}), the fidelity in all the qudits have to be same here. Surprisingly, 
	we find that the symmetry in this model creates hindrances to obtain the optimal refrigeration, i.e., 
	one cannot achieve maximum fidelity in the star network (see Fig. \ref{fig:enter-label} when the measurement outcome is \(P_1^{(2)}\)) and hence the perfect subspace  cooling is not possible. Note also that the suboptimal fidelity can also be obtained when the rank-\(1\) measurement is performed on the central spin. 
	
	\section{Effects of environment in refrigeration}
	\label{sec:evironment_effect}
	
	Environmental influence on all the protocols is inevitable, since the system cannot be kept isolated from its surroundings. We now assume that the regulator is not attached to the bath, but the target systems to be cooled are in contact with the bath.
	
	Let us first consider a scenario where there is a single regulator system and a qudit to be cooled which is attached to a local bath, $E_L$ (see Fig. \ref{fig:schematic}) that comprises  of a collection of harmonic oscillator, given by
	\begin{equation}
		H_{E_L}=\sum_{r=0}^{\omega_c}ha_r^\dagger a_r,
	\end{equation}
	where \(a_r^\dagger\)(\(a_r\)) creation (annihilation) operator of mode \(r\) of the oscillator with \([a_r,a_{r'}^\dagger]=\delta(r-r')\) and \(\omega_c\) is the cut-off frequency of the bath. We consider the temperature of the bath to be \(T_{E_1}\) (i.e., \(\beta_{E_1}=1/k_BT_{E_1}\)) and the system-bath interaction with the qudit is given as
	\begin{equation}
		H_{SE_L}=\sum_{r}[S_L^{+}\otimes a_r + S_L^{-}\otimes a_r^\dagger],
	\end{equation}
	where \(S_L^{+}\)(\(S_L^{-}\)) is the raising (lowering) operator of the target qudit, given as \(S_L^{\pm}=S_L^x\pm iS_L^y\). Due to the presence of the bath, the evolution in between two successive measurement is no more unitary and is governed by the local quatum master (LME) equation \cite{Petruccione},
	\begin{equation}
		\dot{\rho}=-i[H_{total},\rho]+\mathcal{L}(\rho).
	\end{equation}
	Here \(\mathcal{L}(.)\) represents the dissipation due to environment, given by 
	\begin{eqnarray}
		\nonumber\mathcal{L}_L(\rho)&=&\sum_{\omega>0}\gamma_\omega\Big[ (1+n_\omega)(\mathcal{A}\rho \mathcal{A}^\dagger-\frac{1}{2}\{\mathcal{A}^\dagger \mathcal{A},\rho\}\\&&\quad\quad\quad +n_\omega(\mathcal{A}^\dagger\rho \mathcal{A}-\frac{1}{2}\{\mathcal{A}\mathcal{A}^\dagger,\rho\}\Big],
	\end{eqnarray}
	where \(\mathcal{A}\) is Lindblad operator corresponding to LME with \(\mathcal{A}=\frac{1}{2} S_L^{-}\) \cite{konar2022}, \(n_\omega=[\exp(\beta_{E_1}\omega)-1]^{-1}\) and \(\gamma_\omega\) defining the system-bath coupling constant with \(\gamma_\omega\ll     h,\,J\).

	Like the unitary evolution, we again assume that the evolution time between two successive measurement is  \(\tau\), i.e., after time \(\tau\), a projective measurement of rank-\(k\) is performed on the regulator and after the post selection operation, the system is again evolved according to LME. It is expected that the presence of a dissipative environment prohibits to achieve the perfect subspace cooling, but variation of the system-bath coupling play a crucial role in achieving the better subspace cooling. 
	
\begin{figure}
		\centering
		\includegraphics[scale=0.38]{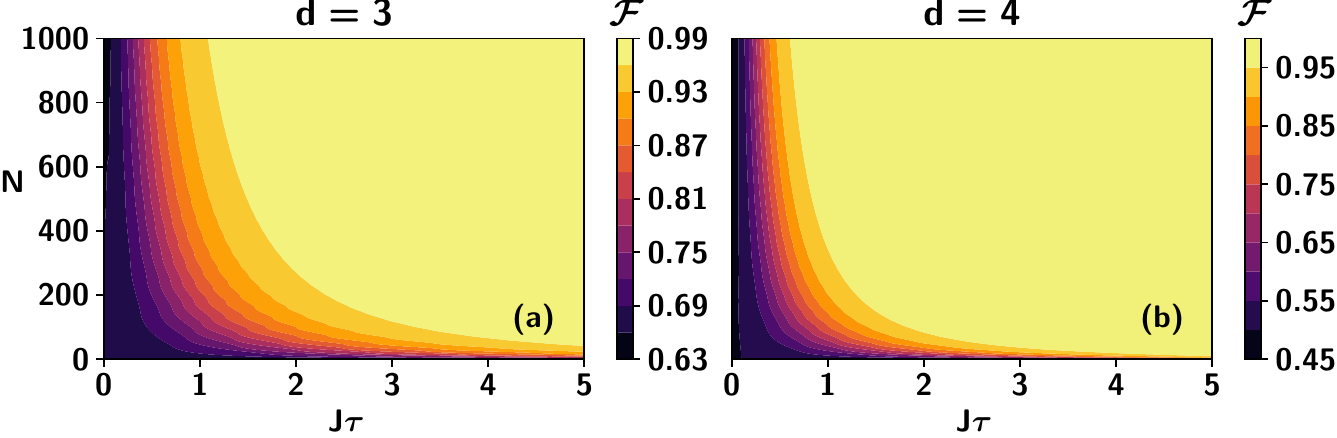}
		\caption{Contour plot of fidelity (\(\mathcal{F}\)) with  \(J\tau\) (\(x\)-axis) and  \(N\) (\(y\)-axis) for different spin dimension, \(d=3\) (a) and \(d=4\) (b). Other parameters  are environment temperature \(T_{E_1}=1\), \(\gamma_\omega=10^{-3}\) and \(\Delta=1\). All other specifications are same as in Fig. \ref{fig:fidelity_N_J}. All axes are dimensionless.}
		\label{fig:open_J_N}
	\end{figure}

	We first observe that in the presence of the bath, one requires much higher number of repetition in evolution and measurements  to achieve the high fidelity compared to the scenario without decoherence. This is true even when the interaction between the regulator and the target system is tuned  suitably, i.e., $J\tau$ is chosen appropriately. Secondly, we  observe that although we require high values of \(N\), a reasonably high fidelity (although not the unit one) can be arrived in the presence of the bath, establishing robustness of this scheme against noise   (see fidelity almost close to unity in the case of \(d=3\) in Fig.  \ref{fig:open_J_N} when  rank-$2$ projector $P_1^{(2)}$ clicks on the regulator and the evolving Hamiltonian is the \(XXZ\) model with \(\Delta =1\)). 
	Thirdly, it decreases with the increase of $d$. Fourthly,  the periodic pattern with \(J\tau\) is lost due to the environmental effect which has a beneficial role in refrigeration. In particular, the  imperfect refrigerating region (i.e., the system parameters for which \(\mathcal{F} <1\)) gets decreased with the increase of spin dimension, thereby demonstrating a dimensional advantage.

	\section{Conclusion}
	\label{sec:conclu}
	In the modern era of quantum technologies, higher-dimensional systems aka qudits can be more beneficial for several quantum information processing tasks ranging from quantum key distribution to implementation of quantum gates in quantum circuits compared with qubits. Further, multiqudit systems have been utilized to demonstrate quantum protocols in experiments.
	

	Starting with a collection of maximally mixed states representing thermal states with infinite temperature, we proposed a method for cooling these states into low-lying energy eigenstates of a local Hamiltonian using a regulatory (auxiliary) qudit, repeated unitary operations among them, and repeated projective measurements with arbitrary ranks on the regulatory qudit. The method can also be called subspace cooling or locking process to low-lying states.
	To accomplish this purpose, the auxiliary system must be constructed in the same state as the target systems to be cooled, and then it interacts with the other target qudits to drive them into a lower energy sector of the local Hamiltonian. Notice that the method presented here cannot be used to cool qubit system in low lying state as only rank-1 projective measurement is possible in two-dimension. In higher dimensions, the rank of the projectors are always chosen to be much less than the dimension of the individual subsystems. 
	
	To demonstrate the effectiveness of the procedure,  we identified several paradigmatic Hamiltonian, namely the one-dimensional spin-s $XXZ$ model including the $XX$ model, and spin-$s$ bilinear and biquadratic Heisenberg (BBH) Hamiltonian in the presence of local magnetic field.  The entire system can be evolved using these Hamiltonians, followed by measurements, which are repeated multiple times, allowing the target system to be locked to low-energy sectors. In particular, we showed that the protocol leads to  unit fidelity  of individual qudits in any dimension provided appropriate  system parameters like interaction strength, time of the evolution, the number of repeated measurements and the rank of the projector for a particular local Hilbert space dimension are chosen. The probability of success for this protocol grows as the rank of the projector increases. Furthermore, we discovered that this procedure is quite sensitive to the interaction pattern between the target and the regulator. In particular, we noticed that when the auxiliary qudit is surrounded by the target qudits, thereby showcasing the star-type network, we can never obtain the unit fidelity, which also diminishes with an increase in dimension. We also extended our studies when the target systems are attached to an environment which is a collection of harmonic oscillator. Although the fidelity gets decreased in the presence of a bath,  we observed that it is not substantial, thereby ensuring its robustness against decoherence. Moreover, we demonstrated that in the presence of an environment, the system parameter dependence can be wiped off. 
	Since the absolute zero temperature cannot be achieved in experiments, the notion of subspace cooling without directly probing the systems may have significant consequences  in laboratory set-up.

	\acknowledgements
	
	We  acknowledge the use of \href{https://github.com/titaschanda/QIClib}{QIClib} -- a modern C++ library for general purpose quantum information processing and quantum computing (\url{https://titaschanda.github.io/QIClib}) and cluster computing facility at Harish-Chandra Research Institute.
	
	\bibliography{ref.bib}

\begin{thebibliography}{92}%
\makeatletter
\providecommand \@ifxundefined [1]{%
 \@ifx{#1\undefined}
}%
\providecommand \@ifnum [1]{%
 \ifnum #1\expandafter \@firstoftwo
 \else \expandafter \@secondoftwo
 \fi
}%
\providecommand \@ifx [1]{%
 \ifx #1\expandafter \@firstoftwo
 \else \expandafter \@secondoftwo
 \fi
}%
\providecommand \natexlab [1]{#1}%
\providecommand \enquote  [1]{``#1''}%
\providecommand \bibnamefont  [1]{#1}%
\providecommand \bibfnamefont [1]{#1}%
\providecommand \citenamefont [1]{#1}%
\providecommand \href@noop [0]{\@secondoftwo}%
\providecommand \href [0]{\begingroup \@sanitize@url \@href}%
\providecommand \@href[1]{\@@startlink{#1}\@@href}%
\providecommand \@@href[1]{\endgroup#1\@@endlink}%
\providecommand \@sanitize@url [0]{\catcode `\\12\catcode `\$12\catcode
  `\&12\catcode `\#12\catcode `\^12\catcode `\_12\catcode `\%12\relax}%
\providecommand \@@startlink[1]{}%
\providecommand \@@endlink[0]{}%
\providecommand \url  [0]{\begingroup\@sanitize@url \@url }%
\providecommand \@url [1]{\endgroup\@href {#1}{\urlprefix }}%
\providecommand \urlprefix  [0]{URL }%
\providecommand \Eprint [0]{\href }%
\providecommand \doibase [0]{http://dx.doi.org/}%
\providecommand \selectlanguage [0]{\@gobble}%
\providecommand \bibinfo  [0]{\@secondoftwo}%
\providecommand \bibfield  [0]{\@secondoftwo}%
\providecommand \translation [1]{[#1]}%
\providecommand \BibitemOpen [0]{}%
\providecommand \bibitemStop [0]{}%
\providecommand \bibitemNoStop [0]{.\EOS\space}%
\providecommand \EOS [0]{\spacefactor3000\relax}%
\providecommand \BibitemShut  [1]{\csname bibitem#1\endcsname}%
\let\auto@bib@innerbib\@empty
\bibitem [{\citenamefont {Gemmer}\ \emph {et~al.}(2004)\citenamefont {Gemmer},
  \citenamefont {Michel},\ and\ \citenamefont {Mahler}}]{gemmer2004}%
  \BibitemOpen
  \bibfield  {author} {\bibinfo {author} {\bibfnamefont {G.}~\bibnamefont
  {Gemmer}}, \bibinfo {author} {\bibfnamefont {M.}~\bibnamefont {Michel}}, \
  and\ \bibinfo {author} {\bibfnamefont {G.}~\bibnamefont {Mahler}},\
  }\href@noop {} {\emph {\bibinfo {title} {Quantum Thermodynamics}}}\ (\bibinfo
   {publisher} {Springer, New York},\ \bibinfo {year} {2004})\BibitemShut
  {NoStop}%
\bibitem [{\citenamefont {Alicki}\ and\ \citenamefont {Fannes}(2013)}]{Alicki}%
  \BibitemOpen
  \bibfield  {author} {\bibinfo {author} {\bibfnamefont {R.}~\bibnamefont
  {Alicki}}\ and\ \bibinfo {author} {\bibfnamefont {M.}~\bibnamefont
  {Fannes}},\ }\href {\doibase 10.1103/PhysRevE.87.042123} {\bibfield
  {journal} {\bibinfo  {journal} {Phys. Rev. E}\ }\textbf {\bibinfo {volume}
  {87}},\ \bibinfo {pages} {042123} (\bibinfo {year} {2013})}\BibitemShut
  {NoStop}%
\bibitem [{\citenamefont {Le}\ \emph {et~al.}(2018)\citenamefont {Le},
  \citenamefont {Levinsen}, \citenamefont {Modi}, \citenamefont {Parish},\ and\
  \citenamefont {Pollock}}]{modi2018}%
  \BibitemOpen
  \bibfield  {author} {\bibinfo {author} {\bibfnamefont {T.~P.}\ \bibnamefont
  {Le}}, \bibinfo {author} {\bibfnamefont {J.}~\bibnamefont {Levinsen}},
  \bibinfo {author} {\bibfnamefont {K.}~\bibnamefont {Modi}}, \bibinfo {author}
  {\bibfnamefont {M.~M.}\ \bibnamefont {Parish}}, \ and\ \bibinfo {author}
  {\bibfnamefont {F.~A.}\ \bibnamefont {Pollock}},\ }\href {\doibase
  10.1103/PhysRevA.97.022106} {\bibfield  {journal} {\bibinfo  {journal} {Phys.
  Rev. A}\ }\textbf {\bibinfo {volume} {97}},\ \bibinfo {pages} {022106}
  (\bibinfo {year} {2018})}\BibitemShut {NoStop}%
\bibitem [{\citenamefont {Ferraro}\ \emph {et~al.}(2018)\citenamefont
  {Ferraro}, \citenamefont {Campisi}, \citenamefont {Andolina}, \citenamefont
  {Pellegrini},\ and\ \citenamefont {Polini}}]{PoliniPRL}%
  \BibitemOpen
  \bibfield  {author} {\bibinfo {author} {\bibfnamefont {D.}~\bibnamefont
  {Ferraro}}, \bibinfo {author} {\bibfnamefont {M.}~\bibnamefont {Campisi}},
  \bibinfo {author} {\bibfnamefont {G.~M.}\ \bibnamefont {Andolina}}, \bibinfo
  {author} {\bibfnamefont {V.}~\bibnamefont {Pellegrini}}, \ and\ \bibinfo
  {author} {\bibfnamefont {M.}~\bibnamefont {Polini}},\ }\href {\doibase
  10.1103/PhysRevLett.120.117702} {\bibfield  {journal} {\bibinfo  {journal}
  {Phys. Rev. Lett.}\ }\textbf {\bibinfo {volume} {120}},\ \bibinfo {pages}
  {117702} (\bibinfo {year} {2018})}\BibitemShut {NoStop}%
\bibitem [{\citenamefont {Ghosh}\ \emph {et~al.}(2020)\citenamefont {Ghosh},
  \citenamefont {Chanda},\ and\ \citenamefont {Sen(De)}}]{srijon2020}%
  \BibitemOpen
  \bibfield  {author} {\bibinfo {author} {\bibfnamefont {S.}~\bibnamefont
  {Ghosh}}, \bibinfo {author} {\bibfnamefont {T.}~\bibnamefont {Chanda}}, \
  and\ \bibinfo {author} {\bibfnamefont {A.}~\bibnamefont {Sen(De)}},\ }\href
  {\doibase 10.1103/PhysRevA.101.032115} {\bibfield  {journal} {\bibinfo
  {journal} {Phys. Rev. A}\ }\textbf {\bibinfo {volume} {101}},\ \bibinfo
  {pages} {032115} (\bibinfo {year} {2020})}\BibitemShut {NoStop}%
\bibitem [{\citenamefont {Campaioli}\ \emph {et~al.}(2024)\citenamefont
  {Campaioli}, \citenamefont {Gherardini}, \citenamefont {Quach}, \citenamefont
  {Polini},\ and\ \citenamefont {Andolina}}]{battery_review_rmp}%
  \BibitemOpen
  \bibfield  {author} {\bibinfo {author} {\bibfnamefont {F.}~\bibnamefont
  {Campaioli}}, \bibinfo {author} {\bibfnamefont {S.}~\bibnamefont
  {Gherardini}}, \bibinfo {author} {\bibfnamefont {J.~Q.}\ \bibnamefont
  {Quach}}, \bibinfo {author} {\bibfnamefont {M.}~\bibnamefont {Polini}}, \
  and\ \bibinfo {author} {\bibfnamefont {G.~M.}\ \bibnamefont {Andolina}},\
  }\href {\doibase 10.1103/RevModPhys.96.031001} {\bibfield  {journal}
  {\bibinfo  {journal} {Rev. Mod. Phys.}\ }\textbf {\bibinfo {volume} {96}},\
  \bibinfo {pages} {031001} (\bibinfo {year} {2024})}\BibitemShut {NoStop}%
\bibitem [{\citenamefont {Ordonez-Miranda}\ \emph {et~al.}(2017)\citenamefont
  {Ordonez-Miranda}, \citenamefont {Ezzahri},\ and\ \citenamefont
  {Joulain}}]{ordonez2017}%
  \BibitemOpen
  \bibfield  {author} {\bibinfo {author} {\bibfnamefont {J.}~\bibnamefont
  {Ordonez-Miranda}}, \bibinfo {author} {\bibfnamefont {Y.}~\bibnamefont
  {Ezzahri}}, \ and\ \bibinfo {author} {\bibfnamefont {K.}~\bibnamefont
  {Joulain}},\ }\href {\doibase 10.1103/PhysRevE.95.022128} {\bibfield
  {journal} {\bibinfo  {journal} {Phys. Rev. E}\ }\textbf {\bibinfo {volume}
  {95}},\ \bibinfo {pages} {022128} (\bibinfo {year} {2017})}\BibitemShut
  {NoStop}%
\bibitem [{\citenamefont {Linden}\ \emph {et~al.}(2010)\citenamefont {Linden},
  \citenamefont {Popescu},\ and\ \citenamefont {Skrzypczyk}}]{linden2010}%
  \BibitemOpen
  \bibfield  {author} {\bibinfo {author} {\bibfnamefont {N.}~\bibnamefont
  {Linden}}, \bibinfo {author} {\bibfnamefont {S.}~\bibnamefont {Popescu}}, \
  and\ \bibinfo {author} {\bibfnamefont {P.}~\bibnamefont {Skrzypczyk}},\
  }\href {\doibase 10.1103/PhysRevLett.105.130401} {\bibfield  {journal}
  {\bibinfo  {journal} {Phys. Rev. Lett.}\ }\textbf {\bibinfo {volume} {105}},\
  \bibinfo {pages} {130401} (\bibinfo {year} {2010})}\BibitemShut {NoStop}%
\bibitem [{\citenamefont {Skrzypczyk}\ \emph {et~al.}(2011)\citenamefont
  {Skrzypczyk}, \citenamefont {Brunner}, \citenamefont {Linden},\ and\
  \citenamefont {Popescu}}]{skrzypczyk2011}%
  \BibitemOpen
  \bibfield  {author} {\bibinfo {author} {\bibfnamefont {P.}~\bibnamefont
  {Skrzypczyk}}, \bibinfo {author} {\bibfnamefont {N.}~\bibnamefont {Brunner}},
  \bibinfo {author} {\bibfnamefont {N.}~\bibnamefont {Linden}}, \ and\ \bibinfo
  {author} {\bibfnamefont {S.}~\bibnamefont {Popescu}},\ }\href {\doibase
  10.1088/1751-8113/44/49/492002} {\bibfield  {journal} {\bibinfo  {journal}
  {Journal of Physics A: Mathematical and Theoretical}\ }\textbf {\bibinfo
  {volume} {44}},\ \bibinfo {pages} {492002} (\bibinfo {year}
  {2011})}\BibitemShut {NoStop}%
\bibitem [{\citenamefont {Konar}\ \emph
  {et~al.}(2022{\natexlab{a}})\citenamefont {Konar}, \citenamefont {Ghosh},
  \citenamefont {Pal},\ and\ \citenamefont {Sen(De)}}]{konar2021}%
  \BibitemOpen
  \bibfield  {author} {\bibinfo {author} {\bibfnamefont {T.~K.}\ \bibnamefont
  {Konar}}, \bibinfo {author} {\bibfnamefont {S.}~\bibnamefont {Ghosh}},
  \bibinfo {author} {\bibfnamefont {A.~K.}\ \bibnamefont {Pal}}, \ and\
  \bibinfo {author} {\bibfnamefont {A.}~\bibnamefont {Sen(De)}},\ }\href
  {\doibase 10.1103/PhysRevA.105.022214} {\bibfield  {journal} {\bibinfo
  {journal} {Phys. Rev. A}\ }\textbf {\bibinfo {volume} {105}},\ \bibinfo
  {pages} {022214} (\bibinfo {year} {2022}{\natexlab{a}})}\BibitemShut
  {NoStop}%
\bibitem [{\citenamefont {Joulain}\ \emph {et~al.}(2016)\citenamefont
  {Joulain}, \citenamefont {Drevillon}, \citenamefont {Ezzahri},\ and\
  \citenamefont {Ordonez-Miranda}}]{karl2016}%
  \BibitemOpen
  \bibfield  {author} {\bibinfo {author} {\bibfnamefont {K.}~\bibnamefont
  {Joulain}}, \bibinfo {author} {\bibfnamefont {J.}~\bibnamefont {Drevillon}},
  \bibinfo {author} {\bibfnamefont {Y.}~\bibnamefont {Ezzahri}}, \ and\
  \bibinfo {author} {\bibfnamefont {J.}~\bibnamefont {Ordonez-Miranda}},\
  }\href {\doibase 10.1103/PhysRevLett.116.200601} {\bibfield  {journal}
  {\bibinfo  {journal} {Phys. Rev. Lett.}\ }\textbf {\bibinfo {volume} {116}},\
  \bibinfo {pages} {200601} (\bibinfo {year} {2016})}\BibitemShut {NoStop}%
\bibitem [{\citenamefont {Ghosh}\ \emph {et~al.}(2021)\citenamefont {Ghosh},
  \citenamefont {Ghoshal},\ and\ \citenamefont {Sen}}]{ahana_transistor}%
  \BibitemOpen
  \bibfield  {author} {\bibinfo {author} {\bibfnamefont {R.}~\bibnamefont
  {Ghosh}}, \bibinfo {author} {\bibfnamefont {A.}~\bibnamefont {Ghoshal}}, \
  and\ \bibinfo {author} {\bibfnamefont {U.}~\bibnamefont {Sen}},\ }\href
  {\doibase 10.1103/PhysRevA.103.052613} {\bibfield  {journal} {\bibinfo
  {journal} {Phys. Rev. A}\ }\textbf {\bibinfo {volume} {103}},\ \bibinfo
  {pages} {052613} (\bibinfo {year} {2021})}\BibitemShut {NoStop}%
\bibitem [{\citenamefont {Scovil}\ and\ \citenamefont
  {Schulz-DuBois}(1959)}]{scovil_prl_heatengine}%
  \BibitemOpen
  \bibfield  {author} {\bibinfo {author} {\bibfnamefont {H.~E.~D.}\
  \bibnamefont {Scovil}}\ and\ \bibinfo {author} {\bibfnamefont {E.~O.}\
  \bibnamefont {Schulz-DuBois}},\ }\href {\doibase 10.1103/PhysRevLett.2.262}
  {\bibfield  {journal} {\bibinfo  {journal} {Phys. Rev. Lett.}\ }\textbf
  {\bibinfo {volume} {2}},\ \bibinfo {pages} {262} (\bibinfo {year}
  {1959})}\BibitemShut {NoStop}%
\bibitem [{\citenamefont {Shapiro}(2012)}]{Shapiro2012}%
  \BibitemOpen
  \bibfield  {author} {\bibinfo {author} {\bibfnamefont {B.}~\bibnamefont
  {Shapiro}},\ }\href {\doibase 10.1088/1751-8113/45/14/143001} {\bibfield
  {journal} {\bibinfo  {journal} {Journal of Physics A: Mathematical and
  Theoretical}\ }\textbf {\bibinfo {volume} {45}},\ \bibinfo {pages} {143001}
  (\bibinfo {year} {2012})}\BibitemShut {NoStop}%
\bibitem [{\citenamefont {H{\"a}ffner}\ \emph {et~al.}(2008)\citenamefont
  {H{\"a}ffner}, \citenamefont {Roos},\ and\ \citenamefont
  {Blatt}}]{haffner2008}%
  \BibitemOpen
  \bibfield  {author} {\bibinfo {author} {\bibfnamefont {H.}~\bibnamefont
  {H{\"a}ffner}}, \bibinfo {author} {\bibfnamefont {C.}~\bibnamefont {Roos}}, \
  and\ \bibinfo {author} {\bibfnamefont {R.}~\bibnamefont {Blatt}},\ }\href
  {\doibase https://doi.org/10.1016/j.physrep.2008.09.003} {\bibfield
  {journal} {\bibinfo  {journal} {Physics Reports}\ }\textbf {\bibinfo {volume}
  {469}},\ \bibinfo {pages} {155 } (\bibinfo {year} {2008})}\BibitemShut
  {NoStop}%
\bibitem [{\citenamefont {Ro{\ss}nagel}\ \emph {et~al.}(2016)\citenamefont
  {Ro{\ss}nagel}, \citenamefont {Dawkins}, \citenamefont {Tolazzi},
  \citenamefont {Abah}, \citenamefont {Lutz}, \citenamefont {Schmidt-Kaler},\
  and\ \citenamefont {Singer}}]{rossnage2016}%
  \BibitemOpen
  \bibfield  {author} {\bibinfo {author} {\bibfnamefont {J.}~\bibnamefont
  {Ro{\ss}nagel}}, \bibinfo {author} {\bibfnamefont {S.~T.}\ \bibnamefont
  {Dawkins}}, \bibinfo {author} {\bibfnamefont {K.~N.}\ \bibnamefont
  {Tolazzi}}, \bibinfo {author} {\bibfnamefont {O.}~\bibnamefont {Abah}},
  \bibinfo {author} {\bibfnamefont {E.}~\bibnamefont {Lutz}}, \bibinfo {author}
  {\bibfnamefont {F.}~\bibnamefont {Schmidt-Kaler}}, \ and\ \bibinfo {author}
  {\bibfnamefont {K.}~\bibnamefont {Singer}},\ }\href {\doibase
  10.1126/science.aad6320} {\bibfield  {journal} {\bibinfo  {journal}
  {Science}\ }\textbf {\bibinfo {volume} {352}},\ \bibinfo {pages} {325}
  (\bibinfo {year} {2016})}\BibitemShut {NoStop}%
\bibitem [{\citenamefont {Karimi}\ and\ \citenamefont
  {Pekola}(2016)}]{karimi_prb_2016}%
  \BibitemOpen
  \bibfield  {author} {\bibinfo {author} {\bibfnamefont {B.}~\bibnamefont
  {Karimi}}\ and\ \bibinfo {author} {\bibfnamefont {J.~P.}\ \bibnamefont
  {Pekola}},\ }\href {\doibase 10.1103/PhysRevB.94.184503} {\bibfield
  {journal} {\bibinfo  {journal} {Phys. Rev. B}\ }\textbf {\bibinfo {volume}
  {94}},\ \bibinfo {pages} {184503} (\bibinfo {year} {2016})}\BibitemShut
  {NoStop}%
\bibitem [{\citenamefont {Zhang}\ \emph {et~al.}(2014)\citenamefont {Zhang},
  \citenamefont {Bariani},\ and\ \citenamefont
  {Meystre}}]{zhang_cavity_engine}%
  \BibitemOpen
  \bibfield  {author} {\bibinfo {author} {\bibfnamefont {K.}~\bibnamefont
  {Zhang}}, \bibinfo {author} {\bibfnamefont {F.}~\bibnamefont {Bariani}}, \
  and\ \bibinfo {author} {\bibfnamefont {P.}~\bibnamefont {Meystre}},\ }\href
  {\doibase 10.1103/PhysRevLett.112.150602} {\bibfield  {journal} {\bibinfo
  {journal} {Phys. Rev. Lett.}\ }\textbf {\bibinfo {volume} {112}},\ \bibinfo
  {pages} {150602} (\bibinfo {year} {2014})}\BibitemShut {NoStop}%
\bibitem [{\citenamefont {Hardal}\ and\ \citenamefont
  {M{\ifmmode\ddot{u}\else\"{u}\fi}stecapl{\ifmmode\imath\else\i\fi}o{\ifmmode\breve{g}\else\u{g}\fi}lu}(2015)}]{Hardal2015Aug}%
  \BibitemOpen
  \bibfield  {author} {\bibinfo {author} {\bibfnamefont
  {A.~{\ifmmode\ddot{U}\else\"{U}\fi}.~C.}\ \bibnamefont {Hardal}}\ and\
  \bibinfo {author} {\bibfnamefont {{\ifmmode\ddot{O}\else\"{O}\fi}.~E.}\
  \bibnamefont
  {M{\ifmmode\ddot{u}\else\"{u}\fi}stecapl{\ifmmode\imath\else\i\fi}o{\ifmmode\breve{g}\else\u{g}\fi}lu}},\
  }\href {\doibase 10.1038/srep12953} {\bibfield  {journal} {\bibinfo
  {journal} {Sci. Rep.}\ }\textbf {\bibinfo {volume} {5}},\ \bibinfo {pages}
  {1} (\bibinfo {year} {2015})}\BibitemShut {NoStop}%
\bibitem [{\citenamefont {Peterson}\ \emph {et~al.}(2019)\citenamefont
  {Peterson}, \citenamefont {Batalh\~ao}, \citenamefont {Herrera},
  \citenamefont {Souza}, \citenamefont {Sarthour}, \citenamefont {Oliveira},\
  and\ \citenamefont {Serra}}]{peterson2019}%
  \BibitemOpen
  \bibfield  {author} {\bibinfo {author} {\bibfnamefont {J.~P.~S.}\
  \bibnamefont {Peterson}}, \bibinfo {author} {\bibfnamefont {T.~B.}\
  \bibnamefont {Batalh\~ao}}, \bibinfo {author} {\bibfnamefont
  {M.}~\bibnamefont {Herrera}}, \bibinfo {author} {\bibfnamefont {A.~M.}\
  \bibnamefont {Souza}}, \bibinfo {author} {\bibfnamefont {R.~S.}\ \bibnamefont
  {Sarthour}}, \bibinfo {author} {\bibfnamefont {I.~S.}\ \bibnamefont
  {Oliveira}}, \ and\ \bibinfo {author} {\bibfnamefont {R.~M.}\ \bibnamefont
  {Serra}},\ }\href {\doibase 10.1103/PhysRevLett.123.240601} {\bibfield
  {journal} {\bibinfo  {journal} {Phys. Rev. Lett.}\ }\textbf {\bibinfo
  {volume} {123}},\ \bibinfo {pages} {240601} (\bibinfo {year}
  {2019})}\BibitemShut {NoStop}%
\bibitem [{\citenamefont {Joshi}\ and\ \citenamefont
  {Mahesh}(2022)}]{mahesh_quantumbattery}%
  \BibitemOpen
  \bibfield  {author} {\bibinfo {author} {\bibfnamefont {J.}~\bibnamefont
  {Joshi}}\ and\ \bibinfo {author} {\bibfnamefont {T.~S.}\ \bibnamefont
  {Mahesh}},\ }\href {\doibase 10.1103/PhysRevA.106.042601} {\bibfield
  {journal} {\bibinfo  {journal} {Phys. Rev. A}\ }\textbf {\bibinfo {volume}
  {106}},\ \bibinfo {pages} {042601} (\bibinfo {year} {2022})}\BibitemShut
  {NoStop}%
\bibitem [{\citenamefont {Raussendorf}\ and\ \citenamefont
  {Briegel}(2001)}]{raussendorf2001}%
  \BibitemOpen
  \bibfield  {author} {\bibinfo {author} {\bibfnamefont {R.}~\bibnamefont
  {Raussendorf}}\ and\ \bibinfo {author} {\bibfnamefont {H.~J.}\ \bibnamefont
  {Briegel}},\ }\href {\doibase 10.1103/PhysRevLett.86.5188} {\bibfield
  {journal} {\bibinfo  {journal} {Phys. Rev. Lett.}\ }\textbf {\bibinfo
  {volume} {86}},\ \bibinfo {pages} {5188} (\bibinfo {year}
  {2001})}\BibitemShut {NoStop}%
\bibitem [{\citenamefont {Horodecki}\ \emph {et~al.}(2009)\citenamefont
  {Horodecki}, \citenamefont {Horodecki}, \citenamefont {Horodecki},\ and\
  \citenamefont {Horodecki}}]{horodecki2009}%
  \BibitemOpen
  \bibfield  {author} {\bibinfo {author} {\bibfnamefont {R.}~\bibnamefont
  {Horodecki}}, \bibinfo {author} {\bibfnamefont {P.}~\bibnamefont
  {Horodecki}}, \bibinfo {author} {\bibfnamefont {M.}~\bibnamefont
  {Horodecki}}, \ and\ \bibinfo {author} {\bibfnamefont {K.}~\bibnamefont
  {Horodecki}},\ }\href {\doibase 10.1103/RevModPhys.81.865} {\bibfield
  {journal} {\bibinfo  {journal} {Rev. Mod. Phys.}\ }\textbf {\bibinfo {volume}
  {81}},\ \bibinfo {pages} {865} (\bibinfo {year} {2009})}\BibitemShut
  {NoStop}%
\bibitem [{\citenamefont {Greenberger}\ \emph {et~al.}(1989)\citenamefont
  {Greenberger}, \citenamefont {Horne},\ and\ \citenamefont
  {Zeilinger}}]{Greenberger2007}%
  \BibitemOpen
  \bibfield  {author} {\bibinfo {author} {\bibfnamefont {D.~M.}\ \bibnamefont
  {Greenberger}}, \bibinfo {author} {\bibfnamefont {M.~A.}\ \bibnamefont
  {Horne}}, \ and\ \bibinfo {author} {\bibfnamefont {A.}~\bibnamefont
  {Zeilinger}},\ }\enquote {\bibinfo {title} {Going beyond bell's theorem},}\
  in\ \href {\doibase 10.1007/978-94-017-0849-4_10} {\emph {\bibinfo
  {booktitle} {Bell's Theorem, Quantum Theory and Conceptions of the
  Universe}}},\ \bibinfo {editor} {edited by\ \bibinfo {editor} {\bibfnamefont
  {M.}~\bibnamefont {Kafatos}}}\ (\bibinfo  {publisher} {Springer
  Netherlands},\ \bibinfo {address} {Dordrecht},\ \bibinfo {year} {1989})\ pp.\
  \bibinfo {pages} {69--72}\BibitemShut {NoStop}%
\bibitem [{\citenamefont {D\"ur}\ \emph {et~al.}(2000)\citenamefont {D\"ur},
  \citenamefont {Vidal},\ and\ \citenamefont {Cirac}}]{Durvidal2002}%
  \BibitemOpen
  \bibfield  {author} {\bibinfo {author} {\bibfnamefont {W.}~\bibnamefont
  {D\"ur}}, \bibinfo {author} {\bibfnamefont {G.}~\bibnamefont {Vidal}}, \ and\
  \bibinfo {author} {\bibfnamefont {J.~I.}\ \bibnamefont {Cirac}},\ }\href
  {\doibase 10.1103/PhysRevA.62.062314} {\bibfield  {journal} {\bibinfo
  {journal} {Phys. Rev. A}\ }\textbf {\bibinfo {volume} {62}},\ \bibinfo
  {pages} {062314} (\bibinfo {year} {2000})}\BibitemShut {NoStop}%
\bibitem [{\citenamefont {Raussendorf}\ \emph {et~al.}(2003)\citenamefont
  {Raussendorf}, \citenamefont {Browne},\ and\ \citenamefont
  {Briegel}}]{mbqc_Raussendorf2003}%
  \BibitemOpen
  \bibfield  {author} {\bibinfo {author} {\bibfnamefont {R.}~\bibnamefont
  {Raussendorf}}, \bibinfo {author} {\bibfnamefont {D.~E.}\ \bibnamefont
  {Browne}}, \ and\ \bibinfo {author} {\bibfnamefont {H.~J.}\ \bibnamefont
  {Briegel}},\ }\href {\doibase 10.1103/PhysRevA.68.022312} {\bibfield
  {journal} {\bibinfo  {journal} {Phys. Rev. A}\ }\textbf {\bibinfo {volume}
  {68}},\ \bibinfo {pages} {022312} (\bibinfo {year} {2003})}\BibitemShut
  {NoStop}%
\bibitem [{\citenamefont {Jozsa}(2005)}]{mbqc1}%
  \BibitemOpen
  \bibfield  {author} {\bibinfo {author} {\bibfnamefont {R.}~\bibnamefont
  {Jozsa}},\ }\href@noop {} {\enquote {\bibinfo {title} {An introduction to
  measurement based quantum computation},}\ } (\bibinfo {year} {2005}),\
  \Eprint {http://arxiv.org/abs/arXiv:quant-ph/0508124}
  {arXiv:quant-ph/0508124} \BibitemShut {NoStop}%
\bibitem [{\citenamefont {Lanyon}\ \emph {et~al.}(2013)\citenamefont {Lanyon},
  \citenamefont {Jurcevic}, \citenamefont {Zwerger}, \citenamefont {Hempel},
  \citenamefont {Martinez}, \citenamefont {D\"ur}, \citenamefont {Briegel},
  \citenamefont {Blatt},\ and\ \citenamefont {Roos}}]{mbqc2}%
  \BibitemOpen
  \bibfield  {author} {\bibinfo {author} {\bibfnamefont {B.~P.}\ \bibnamefont
  {Lanyon}}, \bibinfo {author} {\bibfnamefont {P.}~\bibnamefont {Jurcevic}},
  \bibinfo {author} {\bibfnamefont {M.}~\bibnamefont {Zwerger}}, \bibinfo
  {author} {\bibfnamefont {C.}~\bibnamefont {Hempel}}, \bibinfo {author}
  {\bibfnamefont {E.~A.}\ \bibnamefont {Martinez}}, \bibinfo {author}
  {\bibfnamefont {W.}~\bibnamefont {D\"ur}}, \bibinfo {author} {\bibfnamefont
  {H.~J.}\ \bibnamefont {Briegel}}, \bibinfo {author} {\bibfnamefont
  {R.}~\bibnamefont {Blatt}}, \ and\ \bibinfo {author} {\bibfnamefont {C.~F.}\
  \bibnamefont {Roos}},\ }\href {\doibase 10.1103/PhysRevLett.111.210501}
  {\bibfield  {journal} {\bibinfo  {journal} {Phys. Rev. Lett.}\ }\textbf
  {\bibinfo {volume} {111}},\ \bibinfo {pages} {210501} (\bibinfo {year}
  {2013})}\BibitemShut {NoStop}%
\bibitem [{\citenamefont {Breuer}\ and\ \citenamefont
  {Petruccione}(2007)}]{Petruccione}%
  \BibitemOpen
  \bibfield  {author} {\bibinfo {author} {\bibfnamefont {H.-P.}\ \bibnamefont
  {Breuer}}\ and\ \bibinfo {author} {\bibfnamefont {F.}~\bibnamefont
  {Petruccione}},\ }\href {\doibase 10.1093/acprof:oso/9780199213900.001.0001}
  {\emph {\bibinfo {title} {The Theory of Open Quantum Systems}}}\ (\bibinfo
  {publisher} {Oxford University Press},\ \bibinfo {year} {2007})\BibitemShut
  {NoStop}%
\bibitem [{\citenamefont {Bennett}\ \emph {et~al.}(1996)\citenamefont
  {Bennett}, \citenamefont {Brassard}, \citenamefont {Popescu}, \citenamefont
  {Schumacher}, \citenamefont {Smolin},\ and\ \citenamefont
  {Wootters}}]{Bennett96}%
  \BibitemOpen
  \bibfield  {author} {\bibinfo {author} {\bibfnamefont {C.~H.}\ \bibnamefont
  {Bennett}}, \bibinfo {author} {\bibfnamefont {G.}~\bibnamefont {Brassard}},
  \bibinfo {author} {\bibfnamefont {S.}~\bibnamefont {Popescu}}, \bibinfo
  {author} {\bibfnamefont {B.}~\bibnamefont {Schumacher}}, \bibinfo {author}
  {\bibfnamefont {J.~A.}\ \bibnamefont {Smolin}}, \ and\ \bibinfo {author}
  {\bibfnamefont {W.~K.}\ \bibnamefont {Wootters}},\ }\href {\doibase
  10.1103/PhysRevLett.76.722} {\bibfield  {journal} {\bibinfo  {journal} {Phys.
  Rev. Lett.}\ }\textbf {\bibinfo {volume} {76}},\ \bibinfo {pages} {722}
  (\bibinfo {year} {1996})}\BibitemShut {NoStop}%
\bibitem [{\citenamefont {Nakazato}\ \emph {et~al.}(2003)\citenamefont
  {Nakazato}, \citenamefont {Takazawa},\ and\ \citenamefont
  {Yuasa}}]{nakazato2003}%
  \BibitemOpen
  \bibfield  {author} {\bibinfo {author} {\bibfnamefont {H.}~\bibnamefont
  {Nakazato}}, \bibinfo {author} {\bibfnamefont {T.}~\bibnamefont {Takazawa}},
  \ and\ \bibinfo {author} {\bibfnamefont {K.}~\bibnamefont {Yuasa}},\ }\href
  {\doibase 10.1103/PhysRevLett.90.060401} {\bibfield  {journal} {\bibinfo
  {journal} {Phys. Rev. Lett.}\ }\textbf {\bibinfo {volume} {90}},\ \bibinfo
  {pages} {060401} (\bibinfo {year} {2003})}\BibitemShut {NoStop}%
\bibitem [{\citenamefont {Nakazato}\ \emph {et~al.}(2004)\citenamefont
  {Nakazato}, \citenamefont {Unoki},\ and\ \citenamefont
  {Yuasa}}]{nakazato2004}%
  \BibitemOpen
  \bibfield  {author} {\bibinfo {author} {\bibfnamefont {H.}~\bibnamefont
  {Nakazato}}, \bibinfo {author} {\bibfnamefont {M.}~\bibnamefont {Unoki}}, \
  and\ \bibinfo {author} {\bibfnamefont {K.}~\bibnamefont {Yuasa}},\ }\href
  {\doibase 10.1103/PhysRevA.70.012303} {\bibfield  {journal} {\bibinfo
  {journal} {Phys. Rev. A}\ }\textbf {\bibinfo {volume} {70}},\ \bibinfo
  {pages} {012303} (\bibinfo {year} {2004})}\BibitemShut {NoStop}%
\bibitem [{\citenamefont {Nakazato}\ \emph {et~al.}(2008)\citenamefont
  {Nakazato}, \citenamefont {Yuasa}, \citenamefont {Militello},\ and\
  \citenamefont {Messina}}]{nakazato2008}%
  \BibitemOpen
  \bibfield  {author} {\bibinfo {author} {\bibfnamefont {H.}~\bibnamefont
  {Nakazato}}, \bibinfo {author} {\bibfnamefont {K.}~\bibnamefont {Yuasa}},
  \bibinfo {author} {\bibfnamefont {B.}~\bibnamefont {Militello}}, \ and\
  \bibinfo {author} {\bibfnamefont {A.}~\bibnamefont {Messina}},\ }\href
  {\doibase 10.1103/PhysRevA.77.042114} {\bibfield  {journal} {\bibinfo
  {journal} {Phys. Rev. A}\ }\textbf {\bibinfo {volume} {77}},\ \bibinfo
  {pages} {042114} (\bibinfo {year} {2008})}\BibitemShut {NoStop}%
\bibitem [{\citenamefont {Bellomo}\ \emph {et~al.}(2010)\citenamefont
  {Bellomo}, \citenamefont {Compagno}, \citenamefont {Nakazato},\ and\
  \citenamefont {Yuasa}}]{bellomo2010}%
  \BibitemOpen
  \bibfield  {author} {\bibinfo {author} {\bibfnamefont {B.}~\bibnamefont
  {Bellomo}}, \bibinfo {author} {\bibfnamefont {G.}~\bibnamefont {Compagno}},
  \bibinfo {author} {\bibfnamefont {H.}~\bibnamefont {Nakazato}}, \ and\
  \bibinfo {author} {\bibfnamefont {K.}~\bibnamefont {Yuasa}},\ }\href
  {\doibase 10.1103/PhysRevA.82.060101} {\bibfield  {journal} {\bibinfo
  {journal} {Phys. Rev. A}\ }\textbf {\bibinfo {volume} {82}},\ \bibinfo
  {pages} {060101} (\bibinfo {year} {2010})}\BibitemShut {NoStop}%
\bibitem [{\citenamefont {Burgarth}\ and\ \citenamefont
  {Giovannetti}(2007)}]{burgarth2007}%
  \BibitemOpen
  \bibfield  {author} {\bibinfo {author} {\bibfnamefont {D.}~\bibnamefont
  {Burgarth}}\ and\ \bibinfo {author} {\bibfnamefont {V.}~\bibnamefont
  {Giovannetti}},\ }\href {\doibase 10.1103/PhysRevA.76.062307} {\bibfield
  {journal} {\bibinfo  {journal} {Phys. Rev. A}\ }\textbf {\bibinfo {volume}
  {76}},\ \bibinfo {pages} {062307} (\bibinfo {year} {2007})}\BibitemShut
  {NoStop}%
\bibitem [{\citenamefont {Burgarth}\ and\ \citenamefont
  {Giovannetti}(2008)}]{burgarth2008}%
  \BibitemOpen
  \bibfield  {author} {\bibinfo {author} {\bibfnamefont {D.}~\bibnamefont
  {Burgarth}}\ and\ \bibinfo {author} {\bibfnamefont {V.}~\bibnamefont
  {Giovannetti}},\ }\href {https://doi.org/10.48550/arXiv.0710.0302} {\bibfield
   {journal} {\bibinfo  {journal} {arXiv.0710.0302}\ } (\bibinfo {year}
  {2008})}\BibitemShut {NoStop}%
\bibitem [{\citenamefont {B.~Misra}(1977)}]{misra1977zeno}%
  \BibitemOpen
  \bibfield  {author} {\bibinfo {author} {\bibfnamefont {E.~C. G.~S.}\
  \bibnamefont {B.~Misra}},\ }\href@noop {} {\bibfield  {journal} {\bibinfo
  {journal} {Journal of Mathematical Physics}\ }\textbf {\bibinfo {volume}
  {18}},\ \bibinfo {pages} {756} (\bibinfo {year} {1977})}\BibitemShut
  {NoStop}%
\bibitem [{\citenamefont {Degasperis}\ \emph {et~al.}(1974)\citenamefont
  {Degasperis}, \citenamefont {Fonda},\ and\ \citenamefont
  {Ghirardi}}]{dega1974}%
  \BibitemOpen
  \bibfield  {author} {\bibinfo {author} {\bibfnamefont {A.}~\bibnamefont
  {Degasperis}}, \bibinfo {author} {\bibfnamefont {L.}~\bibnamefont {Fonda}}, \
  and\ \bibinfo {author} {\bibfnamefont {G.~C.}\ \bibnamefont {Ghirardi}},\
  }\href {\doibase 10.1007/BF02731351} {\bibfield  {journal} {\bibinfo
  {journal} {Il Nuovo Cimento A (1965-1970)}\ }\textbf {\bibinfo {volume}
  {21}},\ \bibinfo {pages} {471} (\bibinfo {year} {1974})}\BibitemShut
  {NoStop}%
\bibitem [{\citenamefont {Yan}\ and\ \citenamefont {Jing}(2021)}]{yan2021}%
  \BibitemOpen
  \bibfield  {author} {\bibinfo {author} {\bibfnamefont {J.-s.}\ \bibnamefont
  {Yan}}\ and\ \bibinfo {author} {\bibfnamefont {J.}~\bibnamefont {Jing}},\
  }\href {\doibase 10.1103/PhysRevA.104.105} {\bibfield  {journal} {\bibinfo
  {journal} {Phys. Rev. A}\ }\textbf {\bibinfo {volume} {104}},\ \bibinfo
  {pages} {105} (\bibinfo {year} {2021})}\BibitemShut {NoStop}%
\bibitem [{\citenamefont {Yan}\ and\ \citenamefont {Jing}(2022)}]{yan2022}%
  \BibitemOpen
  \bibfield  {author} {\bibinfo {author} {\bibfnamefont {J.-s.}\ \bibnamefont
  {Yan}}\ and\ \bibinfo {author} {\bibfnamefont {J.}~\bibnamefont {Jing}},\
  }\href {\doibase 10.1103/PhysRevA.105.052607} {\bibfield  {journal} {\bibinfo
   {journal} {Phys. Rev. A}\ }\textbf {\bibinfo {volume} {105}},\ \bibinfo
  {pages} {052607} (\bibinfo {year} {2022})}\BibitemShut {NoStop}%
\bibitem [{\citenamefont {Konar}\ \emph
  {et~al.}(2022{\natexlab{b}})\citenamefont {Konar}, \citenamefont {Ghosh},\
  and\ \citenamefont {Sen(De)}}]{konar_measure_2022}%
  \BibitemOpen
  \bibfield  {author} {\bibinfo {author} {\bibfnamefont {T.~K.}\ \bibnamefont
  {Konar}}, \bibinfo {author} {\bibfnamefont {S.}~\bibnamefont {Ghosh}}, \ and\
  \bibinfo {author} {\bibfnamefont {A.}~\bibnamefont {Sen(De)}},\ }\href
  {\doibase 10.1103/PhysRevA.106.022616} {\bibfield  {journal} {\bibinfo
  {journal} {Phys. Rev. A}\ }\textbf {\bibinfo {volume} {106}},\ \bibinfo
  {pages} {022616} (\bibinfo {year} {2022}{\natexlab{b}})}\BibitemShut
  {NoStop}%
\bibitem [{\citenamefont {Yan}\ and\ \citenamefont
  {Jing}(2023{\natexlab{a}})}]{yan2023}%
  \BibitemOpen
  \bibfield  {author} {\bibinfo {author} {\bibfnamefont {J.-s.}\ \bibnamefont
  {Yan}}\ and\ \bibinfo {author} {\bibfnamefont {J.}~\bibnamefont {Jing}},\
  }\href {\doibase 10.1103/PhysRevA.108.042215} {\bibfield  {journal} {\bibinfo
   {journal} {Phys. Rev. A}\ }\textbf {\bibinfo {volume} {108}},\ \bibinfo
  {pages} {042215} (\bibinfo {year} {2023}{\natexlab{a}})}\BibitemShut
  {NoStop}%
\bibitem [{\citenamefont {Glans}\ \emph {et~al.}(2004)\citenamefont {Glans},
  \citenamefont {Johansson}, \citenamefont {Balasubramanian},\ and\
  \citenamefont {Blake}}]{glans2004}%
  \BibitemOpen
  \bibfield  {author} {\bibinfo {author} {\bibfnamefont {P.-A.}\ \bibnamefont
  {Glans}}, \bibinfo {author} {\bibfnamefont {L.~I.}\ \bibnamefont
  {Johansson}}, \bibinfo {author} {\bibfnamefont {T.}~\bibnamefont
  {Balasubramanian}}, \ and\ \bibinfo {author} {\bibfnamefont {R.~J.}\
  \bibnamefont {Blake}},\ }\href {\doibase 10.1103/PhysRevB.70.033408}
  {\bibfield  {journal} {\bibinfo  {journal} {Phys. Rev. B}\ }\textbf {\bibinfo
  {volume} {70}},\ \bibinfo {pages} {033408} (\bibinfo {year}
  {2004})}\BibitemShut {NoStop}%
\bibitem [{\citenamefont {Romano}(2007)}]{romano2007}%
  \BibitemOpen
  \bibfield  {author} {\bibinfo {author} {\bibfnamefont {R.}~\bibnamefont
  {Romano}},\ }\href {\doibase 10.1103/PhysRevA.75.024301} {\bibfield
  {journal} {\bibinfo  {journal} {Phys. Rev. A}\ }\textbf {\bibinfo {volume}
  {75}},\ \bibinfo {pages} {024301} (\bibinfo {year} {2007})}\BibitemShut
  {NoStop}%
\bibitem [{\citenamefont {Greiner}\ \emph {et~al.}(2017)\citenamefont
  {Greiner}, \citenamefont {Dasari},\ and\ \citenamefont
  {Wrachtrup}}]{Greiner2017Apr}%
  \BibitemOpen
  \bibfield  {author} {\bibinfo {author} {\bibfnamefont {J.~N.}\ \bibnamefont
  {Greiner}}, \bibinfo {author} {\bibfnamefont {D.~B.~R.}\ \bibnamefont
  {Dasari}}, \ and\ \bibinfo {author} {\bibfnamefont {J.}~\bibnamefont
  {Wrachtrup}},\ }\href {\doibase 10.1038/s41598-017-00603-z} {\bibfield
  {journal} {\bibinfo  {journal} {Sci. Rep.}\ }\textbf {\bibinfo {volume}
  {7}},\ \bibinfo {pages} {1} (\bibinfo {year} {2017})}\BibitemShut {NoStop}%
\bibitem [{\citenamefont {Yan}\ and\ \citenamefont
  {Jing}(2023{\natexlab{b}})}]{yan_battery_2023}%
  \BibitemOpen
  \bibfield  {author} {\bibinfo {author} {\bibfnamefont {J.-s.}\ \bibnamefont
  {Yan}}\ and\ \bibinfo {author} {\bibfnamefont {J.}~\bibnamefont {Jing}},\
  }\href {\doibase 10.1103/PhysRevApplied.19.064069} {\bibfield  {journal}
  {\bibinfo  {journal} {Phys. Rev. Appl.}\ }\textbf {\bibinfo {volume} {19}},\
  \bibinfo {pages} {064069} (\bibinfo {year} {2023}{\natexlab{b}})}\BibitemShut
  {NoStop}%
\bibitem [{\citenamefont {Chaki}\ \emph {et~al.}(2023)\citenamefont {Chaki},
  \citenamefont {Bhattacharyya}, \citenamefont {Sen},\ and\ \citenamefont
  {Sen}}]{chaki_battery_2023}%
  \BibitemOpen
  \bibfield  {author} {\bibinfo {author} {\bibfnamefont {P.}~\bibnamefont
  {Chaki}}, \bibinfo {author} {\bibfnamefont {A.}~\bibnamefont
  {Bhattacharyya}}, \bibinfo {author} {\bibfnamefont {K.}~\bibnamefont {Sen}},
  \ and\ \bibinfo {author} {\bibfnamefont {U.}~\bibnamefont {Sen}},\ }\href
  {https://arxiv.org/abs/2307.16856} {\enquote {\bibinfo {title}
  {Auxiliary-assisted stochastic energy extraction from quantum batteries},}\ }
  (\bibinfo {year} {2023}),\ \Eprint {http://arxiv.org/abs/2307.16856}
  {arXiv:2307.16856 [quant-ph]} \BibitemShut {NoStop}%
\bibitem [{\citenamefont {V\'ertesi}\ \emph {et~al.}(2010)\citenamefont
  {V\'ertesi}, \citenamefont {Pironio},\ and\ \citenamefont
  {Brunner}}]{Bellqudits}%
  \BibitemOpen
  \bibfield  {author} {\bibinfo {author} {\bibfnamefont {T.}~\bibnamefont
  {V\'ertesi}}, \bibinfo {author} {\bibfnamefont {S.}~\bibnamefont {Pironio}},
  \ and\ \bibinfo {author} {\bibfnamefont {N.}~\bibnamefont {Brunner}},\ }\href
  {\doibase 10.1103/PhysRevLett.104.060401} {\bibfield  {journal} {\bibinfo
  {journal} {Phys. Rev. Lett.}\ }\textbf {\bibinfo {volume} {104}},\ \bibinfo
  {pages} {060401} (\bibinfo {year} {2010})}\BibitemShut {NoStop}%
\bibitem [{\citenamefont {Brunner}\ \emph {et~al.}(2014)\citenamefont
  {Brunner}, \citenamefont {Cavalcanti}, \citenamefont {Pironio}, \citenamefont
  {Scarani},\ and\ \citenamefont {Wehner}}]{bell-brunner}%
  \BibitemOpen
  \bibfield  {author} {\bibinfo {author} {\bibfnamefont {N.}~\bibnamefont
  {Brunner}}, \bibinfo {author} {\bibfnamefont {D.}~\bibnamefont {Cavalcanti}},
  \bibinfo {author} {\bibfnamefont {S.}~\bibnamefont {Pironio}}, \bibinfo
  {author} {\bibfnamefont {V.}~\bibnamefont {Scarani}}, \ and\ \bibinfo
  {author} {\bibfnamefont {S.}~\bibnamefont {Wehner}},\ }\href {\doibase
  10.1103/RevModPhys.86.419} {\bibfield  {journal} {\bibinfo  {journal} {Rev.
  Mod. Phys.}\ }\textbf {\bibinfo {volume} {86}},\ \bibinfo {pages} {419}
  (\bibinfo {year} {2014})}\BibitemShut {NoStop}%
\bibitem [{\citenamefont {Durt}\ \emph {et~al.}(2004)\citenamefont {Durt},
  \citenamefont {Kaszlikowski}, \citenamefont {Chen},\ and\ \citenamefont
  {Kwek}}]{QKDqudit1}%
  \BibitemOpen
  \bibfield  {author} {\bibinfo {author} {\bibfnamefont {T.}~\bibnamefont
  {Durt}}, \bibinfo {author} {\bibfnamefont {D.}~\bibnamefont {Kaszlikowski}},
  \bibinfo {author} {\bibfnamefont {J.-L.}\ \bibnamefont {Chen}}, \ and\
  \bibinfo {author} {\bibfnamefont {L.~C.}\ \bibnamefont {Kwek}},\ }\href
  {\doibase 10.1103/PhysRevA.69.032313} {\bibfield  {journal} {\bibinfo
  {journal} {Phys. Rev. A}\ }\textbf {\bibinfo {volume} {69}},\ \bibinfo
  {pages} {032313} (\bibinfo {year} {2004})}\BibitemShut {NoStop}%
\bibitem [{\citenamefont {Cerf}\ \emph {et~al.}(2002)\citenamefont {Cerf},
  \citenamefont {Bourennane}, \citenamefont {Karlsson},\ and\ \citenamefont
  {Gisin}}]{gisin2002}%
  \BibitemOpen
  \bibfield  {author} {\bibinfo {author} {\bibfnamefont {N.~J.}\ \bibnamefont
  {Cerf}}, \bibinfo {author} {\bibfnamefont {M.}~\bibnamefont {Bourennane}},
  \bibinfo {author} {\bibfnamefont {A.}~\bibnamefont {Karlsson}}, \ and\
  \bibinfo {author} {\bibfnamefont {N.}~\bibnamefont {Gisin}},\ }\href
  {\doibase 10.1103/PhysRevLett.88.127902} {\bibfield  {journal} {\bibinfo
  {journal} {Phys. Rev. Lett.}\ }\textbf {\bibinfo {volume} {88}},\ \bibinfo
  {pages} {127902} (\bibinfo {year} {2002})}\BibitemShut {NoStop}%
\bibitem [{\citenamefont {Sheridan}\ and\ \citenamefont
  {Scarani}(2010)}]{scarani2010}%
  \BibitemOpen
  \bibfield  {author} {\bibinfo {author} {\bibfnamefont {L.}~\bibnamefont
  {Sheridan}}\ and\ \bibinfo {author} {\bibfnamefont {V.}~\bibnamefont
  {Scarani}},\ }\href {\doibase 10.1103/PhysRevA.82.030301} {\bibfield
  {journal} {\bibinfo  {journal} {Phys. Rev. A}\ }\textbf {\bibinfo {volume}
  {82}},\ \bibinfo {pages} {030301} (\bibinfo {year} {2010})}\BibitemShut
  {NoStop}%
\bibitem [{\citenamefont {Zhou}\ \emph {et~al.}(2003)\citenamefont {Zhou},
  \citenamefont {Zeng}, \citenamefont {Xu},\ and\ \citenamefont
  {Sun}}]{Zhou_2003}%
  \BibitemOpen
  \bibfield  {author} {\bibinfo {author} {\bibfnamefont {D.~L.}\ \bibnamefont
  {Zhou}}, \bibinfo {author} {\bibfnamefont {B.}~\bibnamefont {Zeng}}, \bibinfo
  {author} {\bibfnamefont {Z.}~\bibnamefont {Xu}}, \ and\ \bibinfo {author}
  {\bibfnamefont {C.~P.}\ \bibnamefont {Sun}},\ }\href {\doibase
  10.1103/PhysRevA.68.062303} {\bibfield  {journal} {\bibinfo  {journal} {Phys.
  Rev. A}\ }\textbf {\bibinfo {volume} {68}},\ \bibinfo {pages} {062303}
  (\bibinfo {year} {2003})}\BibitemShut {NoStop}%
\bibitem [{\citenamefont {Hostens}\ \emph {et~al.}(2005)\citenamefont
  {Hostens}, \citenamefont {Dehaene},\ and\ \citenamefont
  {De~Moor}}]{Hosten_2005}%
  \BibitemOpen
  \bibfield  {author} {\bibinfo {author} {\bibfnamefont {E.}~\bibnamefont
  {Hostens}}, \bibinfo {author} {\bibfnamefont {J.}~\bibnamefont {Dehaene}}, \
  and\ \bibinfo {author} {\bibfnamefont {B.}~\bibnamefont {De~Moor}},\ }\href
  {\doibase 10.1103/PhysRevA.71.042315} {\bibfield  {journal} {\bibinfo
  {journal} {Phys. Rev. A}\ }\textbf {\bibinfo {volume} {71}},\ \bibinfo
  {pages} {042315} (\bibinfo {year} {2005})}\BibitemShut {NoStop}%
\bibitem [{\citenamefont {Hall}(2005)}]{hall2005cluster}%
  \BibitemOpen
  \bibfield  {author} {\bibinfo {author} {\bibfnamefont {W.}~\bibnamefont
  {Hall}},\ }\href@noop {} {\bibfield  {journal} {\bibinfo  {journal} {arXiv
  preprint quant-ph/0512130}\ } (\bibinfo {year} {2005})}\BibitemShut {NoStop}%
\bibitem [{\citenamefont {Joo}\ \emph {et~al.}(2007)\citenamefont {Joo},
  \citenamefont {Knight}, \citenamefont {O'Brien},\ and\ \citenamefont
  {Rudolph}}]{joo_2007}%
  \BibitemOpen
  \bibfield  {author} {\bibinfo {author} {\bibfnamefont {J.}~\bibnamefont
  {Joo}}, \bibinfo {author} {\bibfnamefont {P.~L.}\ \bibnamefont {Knight}},
  \bibinfo {author} {\bibfnamefont {J.~L.}\ \bibnamefont {O'Brien}}, \ and\
  \bibinfo {author} {\bibfnamefont {T.}~\bibnamefont {Rudolph}},\ }\href
  {\doibase 10.1103/PhysRevA.76.052326} {\bibfield  {journal} {\bibinfo
  {journal} {Phys. Rev. A}\ }\textbf {\bibinfo {volume} {76}},\ \bibinfo
  {pages} {052326} (\bibinfo {year} {2007})}\BibitemShut {NoStop}%
\bibitem [{\citenamefont {Zhang}\ \emph {et~al.}(2009)\citenamefont {Zhang},
  \citenamefont {Fan},\ and\ \citenamefont {Zhou}}]{Zhou_2009}%
  \BibitemOpen
  \bibfield  {author} {\bibinfo {author} {\bibfnamefont {D.~H.}\ \bibnamefont
  {Zhang}}, \bibinfo {author} {\bibfnamefont {H.}~\bibnamefont {Fan}}, \ and\
  \bibinfo {author} {\bibfnamefont {D.~L.}\ \bibnamefont {Zhou}},\ }\href
  {\doibase 10.1103/PhysRevA.79.042318} {\bibfield  {journal} {\bibinfo
  {journal} {Phys. Rev. A}\ }\textbf {\bibinfo {volume} {79}},\ \bibinfo
  {pages} {042318} (\bibinfo {year} {2009})}\BibitemShut {NoStop}%
\bibitem [{\citenamefont {Wang}\ \emph {et~al.}(2017)\citenamefont {Wang},
  \citenamefont {Stephen},\ and\ \citenamefont {Raussendorf}}]{Raussen_2017}%
  \BibitemOpen
  \bibfield  {author} {\bibinfo {author} {\bibfnamefont {D.-S.}\ \bibnamefont
  {Wang}}, \bibinfo {author} {\bibfnamefont {D.~T.}\ \bibnamefont {Stephen}}, \
  and\ \bibinfo {author} {\bibfnamefont {R.}~\bibnamefont {Raussendorf}},\
  }\href {\doibase 10.1103/PhysRevA.95.032312} {\bibfield  {journal} {\bibinfo
  {journal} {Phys. Rev. A}\ }\textbf {\bibinfo {volume} {95}},\ \bibinfo
  {pages} {032312} (\bibinfo {year} {2017})}\BibitemShut {NoStop}%
\bibitem [{\citenamefont {Wang}\ \emph {et~al.}(2020)\citenamefont {Wang},
  \citenamefont {Hu}, \citenamefont {Sanders},\ and\ \citenamefont
  {Kais}}]{Wang2020}%
  \BibitemOpen
  \bibfield  {author} {\bibinfo {author} {\bibfnamefont {Y.}~\bibnamefont
  {Wang}}, \bibinfo {author} {\bibfnamefont {Z.}~\bibnamefont {Hu}}, \bibinfo
  {author} {\bibfnamefont {B.~C.}\ \bibnamefont {Sanders}}, \ and\ \bibinfo
  {author} {\bibfnamefont {S.}~\bibnamefont {Kais}},\ }\href {\doibase
  10.3389/fphy.2020.589504} {\bibfield  {journal} {\bibinfo  {journal}
  {Frontiers in Physics}\ }\textbf {\bibinfo {volume} {8}} (\bibinfo {year}
  {2020}),\ 10.3389/fphy.2020.589504}\BibitemShut {NoStop}%
\bibitem [{\citenamefont {Booth}\ \emph {et~al.}(2023)\citenamefont {Booth},
  \citenamefont {Kissinger}, \citenamefont {Markham}, \citenamefont
  {Meignant},\ and\ \citenamefont {Perdrix}}]{Booth_2023}%
  \BibitemOpen
  \bibfield  {author} {\bibinfo {author} {\bibfnamefont {R.~I.}\ \bibnamefont
  {Booth}}, \bibinfo {author} {\bibfnamefont {A.}~\bibnamefont {Kissinger}},
  \bibinfo {author} {\bibfnamefont {D.}~\bibnamefont {Markham}}, \bibinfo
  {author} {\bibfnamefont {C.}~\bibnamefont {Meignant}}, \ and\ \bibinfo
  {author} {\bibfnamefont {S.}~\bibnamefont {Perdrix}},\ }\href {\doibase
  10.1088/1751-8121/acbace} {\bibfield  {journal} {\bibinfo  {journal} {Journal
  of Physics A: Mathematical and Theoretical}\ }\textbf {\bibinfo {volume}
  {56}},\ \bibinfo {pages} {115303} (\bibinfo {year} {2023})}\BibitemShut
  {NoStop}%
\bibitem [{\citenamefont {Marqversen}\ and\ \citenamefont
  {Zinner}(2023)}]{Marqversen2023}%
  \BibitemOpen
  \bibfield  {author} {\bibinfo {author} {\bibfnamefont {F.~K.}\ \bibnamefont
  {Marqversen}}\ and\ \bibinfo {author} {\bibfnamefont {N.~T.}\ \bibnamefont
  {Zinner}},\ }\href {\doibase 10.1088/2058-9565/ace2e6} {\bibfield  {journal}
  {\bibinfo  {journal} {Quantum Sci. Technol.}\ }\textbf {\bibinfo {volume}
  {8}},\ \bibinfo {pages} {045001} (\bibinfo {year} {2023})}\BibitemShut
  {NoStop}%
\bibitem [{\citenamefont {Imany}\ \emph {et~al.}(2019)\citenamefont {Imany},
  \citenamefont {Jaramillo-Villegas}, \citenamefont {Alshaykh}, \citenamefont
  {Lukens}, \citenamefont {Odele}, \citenamefont {Moore}, \citenamefont
  {Leaird}, \citenamefont {Qi},\ and\ \citenamefont {Weiner}}]{Imany2019}%
  \BibitemOpen
  \bibfield  {author} {\bibinfo {author} {\bibfnamefont {P.}~\bibnamefont
  {Imany}}, \bibinfo {author} {\bibfnamefont {J.~A.}\ \bibnamefont
  {Jaramillo-Villegas}}, \bibinfo {author} {\bibfnamefont {M.~S.}\ \bibnamefont
  {Alshaykh}}, \bibinfo {author} {\bibfnamefont {J.~M.}\ \bibnamefont
  {Lukens}}, \bibinfo {author} {\bibfnamefont {O.~D.}\ \bibnamefont {Odele}},
  \bibinfo {author} {\bibfnamefont {A.~J.}\ \bibnamefont {Moore}}, \bibinfo
  {author} {\bibfnamefont {D.~E.}\ \bibnamefont {Leaird}}, \bibinfo {author}
  {\bibfnamefont {M.}~\bibnamefont {Qi}}, \ and\ \bibinfo {author}
  {\bibfnamefont {A.~M.}\ \bibnamefont {Weiner}},\ }\href {\doibase
  10.1038/s41534-019-0173-8} {\bibfield  {journal} {\bibinfo  {journal} {npj
  Quantum Information}\ }\textbf {\bibinfo {volume} {5}},\ \bibinfo {pages}
  {59} (\bibinfo {year} {2019})}\BibitemShut {NoStop}%
\bibitem [{\citenamefont {Sawant}\ \emph {et~al.}(2020)\citenamefont {Sawant},
  \citenamefont {Blackmore}, \citenamefont {Gregory}, \citenamefont
  {Mur-Petit}, \citenamefont {Jaksch}, \citenamefont {Aldegunde}, \citenamefont
  {Hutson}, \citenamefont {Tarbutt},\ and\ \citenamefont
  {Cornish}}]{Sawant_2020}%
  \BibitemOpen
  \bibfield  {author} {\bibinfo {author} {\bibfnamefont {R.}~\bibnamefont
  {Sawant}}, \bibinfo {author} {\bibfnamefont {J.~A.}\ \bibnamefont
  {Blackmore}}, \bibinfo {author} {\bibfnamefont {P.~D.}\ \bibnamefont
  {Gregory}}, \bibinfo {author} {\bibfnamefont {J.}~\bibnamefont {Mur-Petit}},
  \bibinfo {author} {\bibfnamefont {D.}~\bibnamefont {Jaksch}}, \bibinfo
  {author} {\bibfnamefont {J.}~\bibnamefont {Aldegunde}}, \bibinfo {author}
  {\bibfnamefont {J.~M.}\ \bibnamefont {Hutson}}, \bibinfo {author}
  {\bibfnamefont {M.~R.}\ \bibnamefont {Tarbutt}}, \ and\ \bibinfo {author}
  {\bibfnamefont {S.~L.}\ \bibnamefont {Cornish}},\ }\href {\doibase
  10.1088/1367-2630/ab60f4} {\bibfield  {journal} {\bibinfo  {journal} {New
  Journal of Physics}\ }\textbf {\bibinfo {volume} {22}},\ \bibinfo {pages}
  {013027} (\bibinfo {year} {2020})}\BibitemShut {NoStop}%
\bibitem [{\citenamefont {Low}\ \emph {et~al.}(2020)\citenamefont {Low},
  \citenamefont {White}, \citenamefont {Cox}, \citenamefont {Day},\ and\
  \citenamefont {Senko}}]{Low2020}%
  \BibitemOpen
  \bibfield  {author} {\bibinfo {author} {\bibfnamefont {P.~J.}\ \bibnamefont
  {Low}}, \bibinfo {author} {\bibfnamefont {B.~M.}\ \bibnamefont {White}},
  \bibinfo {author} {\bibfnamefont {A.~A.}\ \bibnamefont {Cox}}, \bibinfo
  {author} {\bibfnamefont {M.~L.}\ \bibnamefont {Day}}, \ and\ \bibinfo
  {author} {\bibfnamefont {C.}~\bibnamefont {Senko}},\ }\href {\doibase
  10.1103/PhysRevResearch.2.033128} {\bibfield  {journal} {\bibinfo  {journal}
  {Physical Review Research}\ }\textbf {\bibinfo {volume} {2}},\ \bibinfo
  {pages} {033128} (\bibinfo {year} {2020})}\BibitemShut {NoStop}%
\bibitem [{\citenamefont {Hrmo}\ \emph {et~al.}(2023)\citenamefont {Hrmo},
  \citenamefont {Wilhelm}, \citenamefont {Gerster}, \citenamefont {van Mourik},
  \citenamefont {Huber}, \citenamefont {Blatt}, \citenamefont {Schindler},
  \citenamefont {Monz},\ and\ \citenamefont {Ringbauer}}]{Hrmo2023}%
  \BibitemOpen
  \bibfield  {author} {\bibinfo {author} {\bibfnamefont {P.}~\bibnamefont
  {Hrmo}}, \bibinfo {author} {\bibfnamefont {B.}~\bibnamefont {Wilhelm}},
  \bibinfo {author} {\bibfnamefont {L.}~\bibnamefont {Gerster}}, \bibinfo
  {author} {\bibfnamefont {M.~W.}\ \bibnamefont {van Mourik}}, \bibinfo
  {author} {\bibfnamefont {M.}~\bibnamefont {Huber}}, \bibinfo {author}
  {\bibfnamefont {R.}~\bibnamefont {Blatt}}, \bibinfo {author} {\bibfnamefont
  {P.}~\bibnamefont {Schindler}}, \bibinfo {author} {\bibfnamefont
  {T.}~\bibnamefont {Monz}}, \ and\ \bibinfo {author} {\bibfnamefont
  {M.}~\bibnamefont {Ringbauer}},\ }\href {\doibase 10.1038/s41467-023-37375-2}
  {\bibfield  {journal} {\bibinfo  {journal} {Nature Communications}\ }\textbf
  {\bibinfo {volume} {14}} (\bibinfo {year} {2023}),\
  10.1038/s41467-023-37375-2}\BibitemShut {NoStop}%
\bibitem [{\citenamefont {Ghosh}\ and\ \citenamefont
  {Sen(De)}(2022)}]{ghosh2021}%
  \BibitemOpen
  \bibfield  {author} {\bibinfo {author} {\bibfnamefont {S.}~\bibnamefont
  {Ghosh}}\ and\ \bibinfo {author} {\bibfnamefont {A.}~\bibnamefont
  {Sen(De)}},\ }\href {\doibase 10.1103/PhysRevA.105.022628} {\bibfield
  {journal} {\bibinfo  {journal} {Phys. Rev. A}\ }\textbf {\bibinfo {volume}
  {105}},\ \bibinfo {pages} {022628} (\bibinfo {year} {2022})}\BibitemShut
  {NoStop}%
\bibitem [{\citenamefont {Correa}(2014)}]{correa2014}%
  \BibitemOpen
  \bibfield  {author} {\bibinfo {author} {\bibfnamefont {L.~A.}\ \bibnamefont
  {Correa}},\ }\href {\doibase 10.1103/PhysRevE.89.042128} {\bibfield
  {journal} {\bibinfo  {journal} {Phys. Rev. E}\ }\textbf {\bibinfo {volume}
  {89}},\ \bibinfo {pages} {042128} (\bibinfo {year} {2014})}\BibitemShut
  {NoStop}%
\bibitem [{\citenamefont {Wang}\ \emph {et~al.}(2015)\citenamefont {Wang},
  \citenamefont {Lai}, \citenamefont {Ye}, \citenamefont {He}, \citenamefont
  {Ma},\ and\ \citenamefont {Liao}}]{wang2015}%
  \BibitemOpen
  \bibfield  {author} {\bibinfo {author} {\bibfnamefont {J.}~\bibnamefont
  {Wang}}, \bibinfo {author} {\bibfnamefont {Y.}~\bibnamefont {Lai}}, \bibinfo
  {author} {\bibfnamefont {Z.}~\bibnamefont {Ye}}, \bibinfo {author}
  {\bibfnamefont {J.}~\bibnamefont {He}}, \bibinfo {author} {\bibfnamefont
  {Y.}~\bibnamefont {Ma}}, \ and\ \bibinfo {author} {\bibfnamefont
  {Q.}~\bibnamefont {Liao}},\ }\href {\doibase 10.1103/PhysRevE.91.050102}
  {\bibfield  {journal} {\bibinfo  {journal} {Phys. Rev. E}\ }\textbf {\bibinfo
  {volume} {91}},\ \bibinfo {pages} {050102} (\bibinfo {year}
  {2015})}\BibitemShut {NoStop}%
\bibitem [{\citenamefont {Silva}\ \emph {et~al.}(2015)\citenamefont {Silva},
  \citenamefont {Skrzypczyk},\ and\ \citenamefont {Brunner}}]{silva2015}%
  \BibitemOpen
  \bibfield  {author} {\bibinfo {author} {\bibfnamefont {R.}~\bibnamefont
  {Silva}}, \bibinfo {author} {\bibfnamefont {P.}~\bibnamefont {Skrzypczyk}}, \
  and\ \bibinfo {author} {\bibfnamefont {N.}~\bibnamefont {Brunner}},\ }\href
  {\doibase 10.1103/PhysRevE.92.012136} {\bibfield  {journal} {\bibinfo
  {journal} {Phys. Rev. E}\ }\textbf {\bibinfo {volume} {92}},\ \bibinfo
  {pages} {012136} (\bibinfo {year} {2015})}\BibitemShut {NoStop}%
\bibitem [{\citenamefont {Silva}\ \emph {et~al.}(2016)\citenamefont {Silva},
  \citenamefont {Manzano}, \citenamefont {Skrzypczyk},\ and\ \citenamefont
  {Brunner}}]{silva2016}%
  \BibitemOpen
  \bibfield  {author} {\bibinfo {author} {\bibfnamefont {R.}~\bibnamefont
  {Silva}}, \bibinfo {author} {\bibfnamefont {G.}~\bibnamefont {Manzano}},
  \bibinfo {author} {\bibfnamefont {P.}~\bibnamefont {Skrzypczyk}}, \ and\
  \bibinfo {author} {\bibfnamefont {N.}~\bibnamefont {Brunner}},\ }\href
  {\doibase 10.1103/PhysRevE.94.032120} {\bibfield  {journal} {\bibinfo
  {journal} {Phys. Rev. E}\ }\textbf {\bibinfo {volume} {94}},\ \bibinfo
  {pages} {032120} (\bibinfo {year} {2016})}\BibitemShut {NoStop}%
\bibitem [{\citenamefont {Usui}\ \emph {et~al.}(2021)\citenamefont {Usui},
  \citenamefont {Niedenzu},\ and\ \citenamefont {Huber}}]{usui2021}%
  \BibitemOpen
  \bibfield  {author} {\bibinfo {author} {\bibfnamefont {A.}~\bibnamefont
  {Usui}}, \bibinfo {author} {\bibfnamefont {W.}~\bibnamefont {Niedenzu}}, \
  and\ \bibinfo {author} {\bibfnamefont {M.}~\bibnamefont {Huber}},\ }\href
  {\doibase 10.1103/PhysRevA.104.042224} {\bibfield  {journal} {\bibinfo
  {journal} {Phys. Rev. A}\ }\textbf {\bibinfo {volume} {104}},\ \bibinfo
  {pages} {042224} (\bibinfo {year} {2021})}\BibitemShut {NoStop}%
\bibitem [{\citenamefont {Konar}\ \emph {et~al.}(2023)\citenamefont {Konar},
  \citenamefont {Ghosh}, \citenamefont {Pal},\ and\ \citenamefont
  {Sen(De)}}]{konar2022}%
  \BibitemOpen
  \bibfield  {author} {\bibinfo {author} {\bibfnamefont {T.~K.}\ \bibnamefont
  {Konar}}, \bibinfo {author} {\bibfnamefont {S.}~\bibnamefont {Ghosh}},
  \bibinfo {author} {\bibfnamefont {A.~K.}\ \bibnamefont {Pal}}, \ and\
  \bibinfo {author} {\bibfnamefont {A.}~\bibnamefont {Sen(De)}},\ }\href
  {\doibase 10.1103/PhysRevA.107.032602} {\bibfield  {journal} {\bibinfo
  {journal} {Phys. Rev. A}\ }\textbf {\bibinfo {volume} {107}},\ \bibinfo
  {pages} {032602} (\bibinfo {year} {2023})}\BibitemShut {NoStop}%
\bibitem [{\citenamefont {Friedenauer}\ \emph {et~al.}(2008)\citenamefont
  {Friedenauer}, \citenamefont {Schmitz}, \citenamefont {Glueckert},
  \citenamefont {Porras},\ and\ \citenamefont {Schaetz}}]{iontrap08}%
  \BibitemOpen
  \bibfield  {author} {\bibinfo {author} {\bibfnamefont {A.}~\bibnamefont
  {Friedenauer}}, \bibinfo {author} {\bibfnamefont {H.}~\bibnamefont
  {Schmitz}}, \bibinfo {author} {\bibfnamefont {J.~T.}\ \bibnamefont
  {Glueckert}}, \bibinfo {author} {\bibfnamefont {D.}~\bibnamefont {Porras}}, \
  and\ \bibinfo {author} {\bibfnamefont {T.}~\bibnamefont {Schaetz}},\ }\href
  {\doibase 10.1038/nphys1032} {\bibfield  {journal} {\bibinfo  {journal}
  {Nature Physics}\ }\textbf {\bibinfo {volume} {4}},\ \bibinfo {pages} {757}
  (\bibinfo {year} {2008})}\BibitemShut {NoStop}%
\bibitem [{\citenamefont {Kim}\ \emph {et~al.}(2009)\citenamefont {Kim},
  \citenamefont {Chang}, \citenamefont {Islam}, \citenamefont {Korenblit},
  \citenamefont {Duan},\ and\ \citenamefont {Monroe}}]{iontrap09}%
  \BibitemOpen
  \bibfield  {author} {\bibinfo {author} {\bibfnamefont {K.}~\bibnamefont
  {Kim}}, \bibinfo {author} {\bibfnamefont {M.-S.}\ \bibnamefont {Chang}},
  \bibinfo {author} {\bibfnamefont {R.}~\bibnamefont {Islam}}, \bibinfo
  {author} {\bibfnamefont {S.}~\bibnamefont {Korenblit}}, \bibinfo {author}
  {\bibfnamefont {L.-M.}\ \bibnamefont {Duan}}, \ and\ \bibinfo {author}
  {\bibfnamefont {C.}~\bibnamefont {Monroe}},\ }\href {\doibase
  10.1103/PhysRevLett.103.120502} {\bibfield  {journal} {\bibinfo  {journal}
  {Phys. Rev. Lett.}\ }\textbf {\bibinfo {volume} {103}},\ \bibinfo {pages}
  {120502} (\bibinfo {year} {2009})}\BibitemShut {NoStop}%
\bibitem [{\citenamefont {Khromova}\ \emph {et~al.}(2012)\citenamefont
  {Khromova}, \citenamefont {Piltz}, \citenamefont {Scharfenberger},
  \citenamefont {Gloger}, \citenamefont {Johanning}, \citenamefont {Var\'on},\
  and\ \citenamefont {Wunderlich}}]{iontrap12}%
  \BibitemOpen
  \bibfield  {author} {\bibinfo {author} {\bibfnamefont {A.}~\bibnamefont
  {Khromova}}, \bibinfo {author} {\bibfnamefont {C.}~\bibnamefont {Piltz}},
  \bibinfo {author} {\bibfnamefont {B.}~\bibnamefont {Scharfenberger}},
  \bibinfo {author} {\bibfnamefont {T.~F.}\ \bibnamefont {Gloger}}, \bibinfo
  {author} {\bibfnamefont {M.}~\bibnamefont {Johanning}}, \bibinfo {author}
  {\bibfnamefont {A.~F.}\ \bibnamefont {Var\'on}}, \ and\ \bibinfo {author}
  {\bibfnamefont {C.}~\bibnamefont {Wunderlich}},\ }\href {\doibase
  10.1103/PhysRevLett.108.220502} {\bibfield  {journal} {\bibinfo  {journal}
  {Phys. Rev. Lett.}\ }\textbf {\bibinfo {volume} {108}},\ \bibinfo {pages}
  {220502} (\bibinfo {year} {2012})}\BibitemShut {NoStop}%
\bibitem [{\citenamefont {Luo}\ \emph {et~al.}(2019)\citenamefont {Luo},
  \citenamefont {Zhong}, \citenamefont {Erhard}, \citenamefont {Wang},
  \citenamefont {Peng}, \citenamefont {Krenn}, \citenamefont {Jiang},
  \citenamefont {Li}, \citenamefont {Liu}, \citenamefont {Lu}, \citenamefont
  {Zeilinger},\ and\ \citenamefont {Pan}}]{pan2019}%
  \BibitemOpen
  \bibfield  {author} {\bibinfo {author} {\bibfnamefont {Y.-H.}\ \bibnamefont
  {Luo}}, \bibinfo {author} {\bibfnamefont {H.-S.}\ \bibnamefont {Zhong}},
  \bibinfo {author} {\bibfnamefont {M.}~\bibnamefont {Erhard}}, \bibinfo
  {author} {\bibfnamefont {X.-L.}\ \bibnamefont {Wang}}, \bibinfo {author}
  {\bibfnamefont {L.-C.}\ \bibnamefont {Peng}}, \bibinfo {author}
  {\bibfnamefont {M.}~\bibnamefont {Krenn}}, \bibinfo {author} {\bibfnamefont
  {X.}~\bibnamefont {Jiang}}, \bibinfo {author} {\bibfnamefont
  {L.}~\bibnamefont {Li}}, \bibinfo {author} {\bibfnamefont {N.-L.}\
  \bibnamefont {Liu}}, \bibinfo {author} {\bibfnamefont {C.-Y.}\ \bibnamefont
  {Lu}}, \bibinfo {author} {\bibfnamefont {A.}~\bibnamefont {Zeilinger}}, \
  and\ \bibinfo {author} {\bibfnamefont {J.-W.}\ \bibnamefont {Pan}},\ }\href
  {\doibase 10.1103/PhysRevLett.123.070505} {\bibfield  {journal} {\bibinfo
  {journal} {Phys. Rev. Lett.}\ }\textbf {\bibinfo {volume} {123}},\ \bibinfo
  {pages} {070505} (\bibinfo {year} {2019})}\BibitemShut {NoStop}%
\bibitem [{\citenamefont {Wu}\ \emph {et~al.}(2021)\citenamefont {Wu},
  \citenamefont {Zhao}, \citenamefont {Luo}, \citenamefont {Zhong},
  \citenamefont {Peng}, \citenamefont {Chen}, \citenamefont {Xue},
  \citenamefont {Li}, \citenamefont {Liu}, \citenamefont {Lu},\ and\
  \citenamefont {Pan}}]{pan2021}%
  \BibitemOpen
  \bibfield  {author} {\bibinfo {author} {\bibfnamefont {D.}~\bibnamefont
  {Wu}}, \bibinfo {author} {\bibfnamefont {Q.}~\bibnamefont {Zhao}}, \bibinfo
  {author} {\bibfnamefont {Y.-H.}\ \bibnamefont {Luo}}, \bibinfo {author}
  {\bibfnamefont {H.-S.}\ \bibnamefont {Zhong}}, \bibinfo {author}
  {\bibfnamefont {L.-C.}\ \bibnamefont {Peng}}, \bibinfo {author}
  {\bibfnamefont {K.}~\bibnamefont {Chen}}, \bibinfo {author} {\bibfnamefont
  {P.}~\bibnamefont {Xue}}, \bibinfo {author} {\bibfnamefont {L.}~\bibnamefont
  {Li}}, \bibinfo {author} {\bibfnamefont {N.-L.}\ \bibnamefont {Liu}},
  \bibinfo {author} {\bibfnamefont {C.-Y.}\ \bibnamefont {Lu}}, \ and\ \bibinfo
  {author} {\bibfnamefont {J.-W.}\ \bibnamefont {Pan}},\ }\href {\doibase
  10.1103/PhysRevResearch.3.023017} {\bibfield  {journal} {\bibinfo  {journal}
  {Phys. Rev. Res.}\ }\textbf {\bibinfo {volume} {3}},\ \bibinfo {pages}
  {023017} (\bibinfo {year} {2021})}\BibitemShut {NoStop}%
\bibitem [{\citenamefont {Rasmussen}\ \emph {et~al.}(2021)\citenamefont
  {Rasmussen}, \citenamefont {Christensen}, \citenamefont {Pedersen},
  \citenamefont {Kristensen}, \citenamefont {B\ae{}kkegaard}, \citenamefont
  {Loft},\ and\ \citenamefont {Zinner}}]{PRXQuantumsuper}%
  \BibitemOpen
  \bibfield  {author} {\bibinfo {author} {\bibfnamefont {S.}~\bibnamefont
  {Rasmussen}}, \bibinfo {author} {\bibfnamefont {K.}~\bibnamefont
  {Christensen}}, \bibinfo {author} {\bibfnamefont {S.}~\bibnamefont
  {Pedersen}}, \bibinfo {author} {\bibfnamefont {L.}~\bibnamefont
  {Kristensen}}, \bibinfo {author} {\bibfnamefont {T.}~\bibnamefont
  {B\ae{}kkegaard}}, \bibinfo {author} {\bibfnamefont {N.}~\bibnamefont
  {Loft}}, \ and\ \bibinfo {author} {\bibfnamefont {N.}~\bibnamefont
  {Zinner}},\ }\href {\doibase 10.1103/PRXQuantum.2.040204} {\bibfield
  {journal} {\bibinfo  {journal} {PRX Quantum}\ }\textbf {\bibinfo {volume}
  {2}},\ \bibinfo {pages} {040204} (\bibinfo {year} {2021})}\BibitemShut
  {NoStop}%
\bibitem [{\citenamefont {Ying}\ \emph {et~al.}(2023)\citenamefont {Ying},
  \citenamefont {Cheng}, \citenamefont {Zhao}, \citenamefont {Huang},
  \citenamefont {Zhang}, \citenamefont {Gong}, \citenamefont {Wu},
  \citenamefont {Wang}, \citenamefont {Liang}, \citenamefont {Lin},
  \citenamefont {Xu}, \citenamefont {Deng}, \citenamefont {Rong}, \citenamefont
  {Peng}, \citenamefont {Yung}, \citenamefont {Zhu},\ and\ \citenamefont
  {Pan}}]{ying23}%
  \BibitemOpen
  \bibfield  {author} {\bibinfo {author} {\bibfnamefont {C.}~\bibnamefont
  {Ying}}, \bibinfo {author} {\bibfnamefont {B.}~\bibnamefont {Cheng}},
  \bibinfo {author} {\bibfnamefont {Y.}~\bibnamefont {Zhao}}, \bibinfo {author}
  {\bibfnamefont {H.-L.}\ \bibnamefont {Huang}}, \bibinfo {author}
  {\bibfnamefont {Y.-N.}\ \bibnamefont {Zhang}}, \bibinfo {author}
  {\bibfnamefont {M.}~\bibnamefont {Gong}}, \bibinfo {author} {\bibfnamefont
  {Y.}~\bibnamefont {Wu}}, \bibinfo {author} {\bibfnamefont {S.}~\bibnamefont
  {Wang}}, \bibinfo {author} {\bibfnamefont {F.}~\bibnamefont {Liang}},
  \bibinfo {author} {\bibfnamefont {J.}~\bibnamefont {Lin}}, \bibinfo {author}
  {\bibfnamefont {Y.}~\bibnamefont {Xu}}, \bibinfo {author} {\bibfnamefont
  {H.}~\bibnamefont {Deng}}, \bibinfo {author} {\bibfnamefont {H.}~\bibnamefont
  {Rong}}, \bibinfo {author} {\bibfnamefont {C.-Z.}\ \bibnamefont {Peng}},
  \bibinfo {author} {\bibfnamefont {M.-H.}\ \bibnamefont {Yung}}, \bibinfo
  {author} {\bibfnamefont {X.}~\bibnamefont {Zhu}}, \ and\ \bibinfo {author}
  {\bibfnamefont {J.-W.}\ \bibnamefont {Pan}},\ }\href {\doibase
  10.1103/PhysRevLett.130.110601} {\bibfield  {journal} {\bibinfo  {journal}
  {Phys. Rev. Lett.}\ }\textbf {\bibinfo {volume} {130}},\ \bibinfo {pages}
  {110601} (\bibinfo {year} {2023})}\BibitemShut {NoStop}%
\bibitem [{\citenamefont {Uhlmann}(1976)}]{uhlmann1976}%
  \BibitemOpen
  \bibfield  {author} {\bibinfo {author} {\bibfnamefont {A.}~\bibnamefont
  {Uhlmann}},\ }\href {\doibase https://doi.org/10.1016/0034-4877(76)90060-4}
  {\bibfield  {journal} {\bibinfo  {journal} {Reports on Mathematical Physics}\
  }\textbf {\bibinfo {volume} {9}},\ \bibinfo {pages} {273} (\bibinfo {year}
  {1976})}\BibitemShut {NoStop}%
\bibitem [{\citenamefont {Jozsa}(1994)}]{jozsa1994}%
  \BibitemOpen
  \bibfield  {author} {\bibinfo {author} {\bibfnamefont {R.}~\bibnamefont
  {Jozsa}},\ }\href {\doibase 10.1080/09500349414552171} {\bibfield  {journal}
  {\bibinfo  {journal} {Journal of Modern Optics}\ }\textbf {\bibinfo {volume}
  {41}},\ \bibinfo {pages} {2315} (\bibinfo {year} {1994})},\ \Eprint
  {http://arxiv.org/abs/https://doi.org/10.1080/09500349414552171}
  {https://doi.org/10.1080/09500349414552171} \BibitemShut {NoStop}%
\bibitem [{\citenamefont {Mendon\ifmmode~\mbox{\c{c}}\else \c{c}\fi{}a}\ \emph
  {et~al.}(2008)\citenamefont {Mendon\ifmmode~\mbox{\c{c}}\else \c{c}\fi{}a},
  \citenamefont {Napolitano}, \citenamefont {Marchiolli}, \citenamefont
  {Foster},\ and\ \citenamefont {Liang}}]{mendon2008}%
  \BibitemOpen
  \bibfield  {author} {\bibinfo {author} {\bibfnamefont {P.~E. M.~F.}\
  \bibnamefont {Mendon\ifmmode~\mbox{\c{c}}\else \c{c}\fi{}a}}, \bibinfo
  {author} {\bibfnamefont {R.~d.~J.}\ \bibnamefont {Napolitano}}, \bibinfo
  {author} {\bibfnamefont {M.~A.}\ \bibnamefont {Marchiolli}}, \bibinfo
  {author} {\bibfnamefont {C.~J.}\ \bibnamefont {Foster}}, \ and\ \bibinfo
  {author} {\bibfnamefont {Y.-C.}\ \bibnamefont {Liang}},\ }\href {\doibase
  10.1103/PhysRevA.78.052330} {\bibfield  {journal} {\bibinfo  {journal} {Phys.
  Rev. A}\ }\textbf {\bibinfo {volume} {78}},\ \bibinfo {pages} {052330}
  (\bibinfo {year} {2008})}\BibitemShut {NoStop}%
\bibitem [{\citenamefont {Miszczak}\ \emph {et~al.}(2008)\citenamefont
  {Miszczak}, \citenamefont {Puchała}, \citenamefont {Horodecki},
  \citenamefont {Uhlmann},\ and\ \citenamefont {Życzkowski}}]{miszczak2008}%
  \BibitemOpen
  \bibfield  {author} {\bibinfo {author} {\bibfnamefont {J.~A.}\ \bibnamefont
  {Miszczak}}, \bibinfo {author} {\bibfnamefont {Z.}~\bibnamefont {Puchała}},
  \bibinfo {author} {\bibfnamefont {P.}~\bibnamefont {Horodecki}}, \bibinfo
  {author} {\bibfnamefont {A.}~\bibnamefont {Uhlmann}}, \ and\ \bibinfo
  {author} {\bibfnamefont {K.}~\bibnamefont {Życzkowski}},\ }\href@noop {}
  {\enquote {\bibinfo {title} {Sub-- and super--fidelity as bounds for quantum
  fidelity},}\ } (\bibinfo {year} {2008}),\ \Eprint
  {http://arxiv.org/abs/0805.2037} {arXiv:0805.2037 [quant-ph]} \BibitemShut
  {NoStop}%
\bibitem [{\citenamefont {Chakrabarti}\ \emph {et~al.}(1996)\citenamefont
  {Chakrabarti}, \citenamefont {Dutta},\ and\ \citenamefont {Sen}}]{dutta}%
  \BibitemOpen
  \bibfield  {author} {\bibinfo {author} {\bibfnamefont {B.~K.}\ \bibnamefont
  {Chakrabarti}}, \bibinfo {author} {\bibfnamefont {A.}~\bibnamefont {Dutta}},
  \ and\ \bibinfo {author} {\bibfnamefont {P.}~\bibnamefont {Sen}},\ }\href
  {https://link.springer.com/book/10.1007/978-3-540-49865-0} {\emph {\bibinfo
  {title} {{Quantum Ising Phases and Transitions in Transverse Ising
  Models}}}}\ (\bibinfo  {publisher} {Springer},\ \bibinfo {address} {Berlin,
  Germany},\ \bibinfo {year} {1996})\BibitemShut {NoStop}%
\bibitem [{\citenamefont {Sutherland}(1975)}]{Sutherland1975}%
  \BibitemOpen
  \bibfield  {author} {\bibinfo {author} {\bibfnamefont {B.}~\bibnamefont
  {Sutherland}},\ }\href {\doibase 10.1103/PhysRevB.12.3795} {\bibfield
  {journal} {\bibinfo  {journal} {Phys. Rev. B}\ }\textbf {\bibinfo {volume}
  {12}},\ \bibinfo {pages} {3795} (\bibinfo {year} {1975})}\BibitemShut
  {NoStop}%
\bibitem [{\citenamefont {Takhtajan}(1982)}]{Takhtajan1982}%
  \BibitemOpen
  \bibfield  {author} {\bibinfo {author} {\bibfnamefont {L.}~\bibnamefont
  {Takhtajan}},\ }\href {\doibase https://doi.org/10.1016/0375-9601(82)90764-2}
  {\bibfield  {journal} {\bibinfo  {journal} {Phys. Lett. A}\ }\textbf
  {\bibinfo {volume} {87}},\ \bibinfo {pages} {479} (\bibinfo {year}
  {1982})}\BibitemShut {NoStop}%
\bibitem [{\citenamefont {Babujian}(1982)}]{Babu82}%
  \BibitemOpen
  \bibfield  {author} {\bibinfo {author} {\bibfnamefont {H.}~\bibnamefont
  {Babujian}},\ }\href {\doibase https://doi.org/10.1016/0375-9601(82)90403-0}
  {\bibfield  {journal} {\bibinfo  {journal} {Phys. Lett. A}\ }\textbf
  {\bibinfo {volume} {90}},\ \bibinfo {pages} {479} (\bibinfo {year}
  {1982})}\BibitemShut {NoStop}%
\bibitem [{\citenamefont {F\'ath}\ and\ \citenamefont
  {S\'olyom}(1991)}]{Fath91}%
  \BibitemOpen
  \bibfield  {author} {\bibinfo {author} {\bibfnamefont {G.}~\bibnamefont
  {F\'ath}}\ and\ \bibinfo {author} {\bibfnamefont {J.}~\bibnamefont
  {S\'olyom}},\ }\href {\doibase 10.1103/PhysRevB.44.11836} {\bibfield
  {journal} {\bibinfo  {journal} {Phys. Rev. B}\ }\textbf {\bibinfo {volume}
  {44}},\ \bibinfo {pages} {11836} (\bibinfo {year} {1991})}\BibitemShut
  {NoStop}%
\bibitem [{\citenamefont {F\'ath}\ and\ \citenamefont
  {S\'olyom}(1993)}]{Fath93}%
  \BibitemOpen
  \bibfield  {author} {\bibinfo {author} {\bibfnamefont {G.}~\bibnamefont
  {F\'ath}}\ and\ \bibinfo {author} {\bibfnamefont {J.}~\bibnamefont
  {S\'olyom}},\ }\href {\doibase 10.1103/PhysRevB.47.872} {\bibfield  {journal}
  {\bibinfo  {journal} {Phys. Rev. B}\ }\textbf {\bibinfo {volume} {47}},\
  \bibinfo {pages} {872} (\bibinfo {year} {1993})}\BibitemShut {NoStop}%
\bibitem [{\citenamefont {Rakov}\ and\ \citenamefont
  {Weyrauch}(2022)}]{BBH2022}%
  \BibitemOpen
  \bibfield  {author} {\bibinfo {author} {\bibfnamefont {M.~V.}\ \bibnamefont
  {Rakov}}\ and\ \bibinfo {author} {\bibfnamefont {M.}~\bibnamefont
  {Weyrauch}},\ }\href {\doibase 10.1103/PhysRevB.105.024424} {\bibfield
  {journal} {\bibinfo  {journal} {Phys. Rev. B}\ }\textbf {\bibinfo {volume}
  {105}},\ \bibinfo {pages} {024424} (\bibinfo {year} {2022})}\BibitemShut
  {NoStop}%
\bibitem [{\citenamefont {Mahesh}\ \emph {et~al.}(2021)\citenamefont {Mahesh},
  \citenamefont {Khurana}, \citenamefont {Krithika}, \citenamefont {Sreejith},\
  and\ \citenamefont {Kumar}}]{Mahesh2021_startopology}%
  \BibitemOpen
  \bibfield  {author} {\bibinfo {author} {\bibfnamefont {T.~S.}\ \bibnamefont
  {Mahesh}}, \bibinfo {author} {\bibfnamefont {D.}~\bibnamefont {Khurana}},
  \bibinfo {author} {\bibfnamefont {V.~R.}\ \bibnamefont {Krithika}}, \bibinfo
  {author} {\bibfnamefont {G.~J.}\ \bibnamefont {Sreejith}}, \ and\ \bibinfo
  {author} {\bibfnamefont {C.~S.~S.}\ \bibnamefont {Kumar}},\ }\href {\doibase
  10.1088/1361-648X/ac0dd3} {\bibfield  {journal} {\bibinfo  {journal} {J.
  Phys.: Condens. Matter}\ }\textbf {\bibinfo {volume} {33}},\ \bibinfo {pages}
  {383002} (\bibinfo {year} {2021})}\BibitemShut {NoStop}%
\bibitem [{\citenamefont {Pande}\ \emph {et~al.}(2017)\citenamefont {Pande},
  \citenamefont {Bhole}, \citenamefont {Khurana},\ and\ \citenamefont
  {Mahesh}}]{mahesh_star_cool}%
  \BibitemOpen
  \bibfield  {author} {\bibinfo {author} {\bibfnamefont {V.~R.}\ \bibnamefont
  {Pande}}, \bibinfo {author} {\bibfnamefont {G.}~\bibnamefont {Bhole}},
  \bibinfo {author} {\bibfnamefont {D.}~\bibnamefont {Khurana}}, \ and\
  \bibinfo {author} {\bibfnamefont {T.~S.}\ \bibnamefont {Mahesh}},\ }\href
  {\doibase 10.1103/PhysRevA.96.012330} {\bibfield  {journal} {\bibinfo
  {journal} {Phys. Rev. A}\ }\textbf {\bibinfo {volume} {96}},\ \bibinfo
  {pages} {012330} (\bibinfo {year} {2017})}\BibitemShut {NoStop}%
\end{thebibliography}%

\end{document}